\documentclass[aps,prc,preprint,amsmath,showpacs,superscriptaddress,nofootinbib]{revtex4}
\usepackage{graphicx,epsfig}
\newcommand{\beqy}{\begin{eqnarray}}
\newcommand{\eeqy}{\end{eqnarray}}
\newcommand{\bmlet}{\begin{subequations}}
\newcommand{\emlet}{\end{subequations}}

\begin{document}

\textwidth 16.2 cm
\oddsidemargin -.54 cm
\evensidemargin -.54 cm

\def\gsimeq{\,\,\raise0.14em\hbox{$>$}\kern-0.76em\lower0.28em\hbox
{$\sim$}\,\,}
\def\lsimeq{\,\,\raise0.14em\hbox{$<$}\kern-0.76em\lower0.28em\hbox
{$\sim$}\,\,}

\title{Further explorations of Skyrme-Hartree-Fock-Bogoliubov mass formulas.
XII: Stiffness and stability of neutron-star matter}
\author{S.~Goriely}
\affiliation{Institut d'Astronomie et d'Astrophysique, CP-226,
Universit\'e Libre de Bruxelles, B-1050 Brussels, Belgium}
\author{N.~Chamel}
\affiliation{Institut d'Astronomie et d'Astrophysique, CP-226, Universit\'e
Libre de Bruxelles, B-1050 Brussels, Belgium}
\author{J.~M.~Pearson}
\affiliation{D\'epartement de Physique, Universit\'e de Montr\'eal, Montr\'eal 
(Qu\'ebec), H3C 3J7 Canada}
\date{\today}

\begin{abstract}
We construct three new Hartree-Fock-Bogoliubov (HFB) mass models, labeled 
HFB-19, HFB-20, and HFB-21, with unconventional Skyrme forces containing $t_4$ 
and $t_5$ terms, i.e., density-dependent generalizations of the usual $t_1$ and
$t_2$ terms, respectively. The new forces underlying these models are fitted
respectively to three different realistic equations of state of neutron matter
for which the density dependence of the symmetry energy ranges from the very 
soft to the very stiff, reflecting thereby our present lack of complete
knowledge of the high-density behavior of nuclear matter. All unphysical 
instabilities of nuclear matter, including the
transition to a polarized state in neutron-star matter, are eliminated with 
the new forces. At the same time the new 
models fit essentially all the available mass data with rms deviations of 
0.58 MeV and give the same high quality fits to measured charge 
radii that we obtained in earlier models with conventional Skyrme forces. 
Being constrained by neutron matter, these new mass models, which all give 
similar extrapolations out to the neutron drip line, are highly
appropriate for studies of the $r$-process and the outer crust of neutron stars.
Moreover, the underlying forces, labeled BSk19, BSk20 and BSk21, respectively, 
are well adapted to the study of the inner crust and core of neutron stars.
The new family of Skyrme forces thus opens the way to a unified
description of all regions of neutron stars. 
\end{abstract}


\maketitle

\section{Introduction}

With a view to their astrophysical application in neutron-rich environments,
we have for some years been developing a series of nuclear-mass models based 
on the Hartree-Fock-Bogoliubov (HFB) method with Skyrme and contact-pairing 
forces, together with phenomenological Wigner terms and correction terms for 
the spurious collective energy; all the model parameters have been fitted to 
essentially all the experimental mass data. The different Skyrme forces used in
each of our models up to and including HFB-17 \cite{gcp09} had the conventional
form
\beqy \label{1}
v_{ij} & = &
t_0(1+x_0 P_\sigma)\delta({\pmb{r}_{ij}})
+\frac{1}{2} t_1(1+x_1 P_\sigma)\frac{1}{\hbar^2}\left[p_{ij}^2\,
\delta({\pmb{r}_{ij}}) +\delta({\pmb{r}_{ij}})\, p_{ij}^2 \right]\nonumber\\
& &+t_2(1+x_2 P_\sigma)\frac{1}{\hbar^2}\pmb{p}_{ij}.\delta(\pmb{r}_{ij})\,
 \pmb{p}_{ij}
+\frac{1}{6}t_3(1+x_3 P_\sigma)\rho(\pmb{r})^\alpha\,\delta(\pmb{r}_{ij})
\nonumber\\
& &+\frac{\rm i}{\hbar^2}W_0(\mbox{\boldmath$\sigma_i+\sigma_j$})\cdot
\pmb{p}_{ij}\times\delta(\pmb{r}_{ij})\,\pmb{p}_{ij}  \quad ,
\eeqy
where $\pmb{r}_{ij} = \pmb{r}_i - \pmb{r}_j$, $\pmb{r} = (\pmb{r}_i +
\pmb{r}_j)/2$, $\pmb{p}_{ij} = - {\rm i}\hbar(\pmb{\nabla}_i-\pmb{\nabla}_j)/2$
is the relative momentum, $P_\sigma$ is the two-body spin-exchange
operator, and $\rho(\pmb{r}) = \rho_n(\pmb{r}) + \rho_p(\pmb{r})$ is the total
local density, $\rho_n(\pmb{r})$ and $\rho_p(\pmb{r})$ being the neutron and
proton densities, respectively. With this model we were able to fit with an rms
deviation of 0.581 MeV the 2149 measured masses of nuclei with $N$ and $Z \ge$ 
8 given in the 2003 Atomic Mass Evaluation~\cite{audi03} (AME), while at the 
same time constraining the underlying Skyrme force to fit properties of 
homogeneous neutron matter (NeuM), as determined by many-body calculations
with realistic two- and three-nucleon forces; the pairing force was fitted to
the $^1S_0$ pairing gaps of homogeneous nuclear matter of the 
appropriate charge asymmetry. 

The constraint of the Skyrme force to NeuM enhances the reliability with which
model HFB-17 can make mass predictions for the experimentally inaccessible
highly neutron-rich nuclei that appear in the outer crust of neutron stars and 
that are involved in the $r$-process of stellar nucleosynthesis. Moreover, such a
mass model can be used to extrapolate beyond the drip line to the inner crust 
of neutron stars, using the underlying force (BSk17 in the case of the HFB-17
mass model) to calculate the equation of state (EOS) in this region, i.e., the
energy per nucleon as a function of density for a given temperature. Our 
confidence in this extrapolation derives not only from the fit of the 
interactions to NeuM but also from the precision fit to masses, 
which means that the presence of protons and the existence of inhomogeneities 
in the inner crust are well represented. Being able to use the same effective 
force for the outer and inner crusts means that a coherent treatment of the 
interface between the two regions becomes possible~\cite{onsi08}.

The fit to the calculated properties of NeuM suggests that the extrapolation 
beyond the drip line can be extended still further, into the homogeneous core 
of neutron stars, making it possible to use the model force to calculate the 
EOS of the matter of that region, the so-called neutron-star matter (N*M),
which is highly neutron rich but contains also proton-electron pairs and 
possibly muons (and other particles, such as hyperons, quarks, etc., 
toward the center of the star). Being able to use the effective force in this
region would provide a valuable tool for extending the information
given directly by the realistic calculations of NeuM to which the force had
been fitted. In particular it could be indispensable for a calculation of the 
EOS of N*M, realistic calculations being extremely limited in this respect,
if not non-existent. In this way we would open up the 
prospect of a unified treatment of all regions of the neutron star 
using a unique effective force; in particular a coherent treatment of the
interface between the inner crust and the core would become possible, as
for the interface between the inner and outer crusts.

However, 
in the case of HFB-17 and all our previous models, this development was 
frustrated by the fact that they predict NeuM and N*M to have polarized
ground states that would lead to neutron stars being unstable against
collapse to indefinitely high densities: the unpolarized state of NeuM to which
our forces were fitted turns out not to be the ground state. 
We dealt with this problem in our latest published model, HFB-18 \cite{cgp09},
by adding to the conventional form of Skyrme force (\ref{1}) two extra terms, 
writing our complete Skyrme force as
\beqy
\label{2}
v^{\prime}_{i,j} &=& v_{i,j} +
\frac{1}{2}\,t_4(1+x_4 P_\sigma)\frac{1}{\hbar^2} \left\{p_{ij}^2\,
\rho({\pmb{r}})^\beta\,\delta({\pmb{r}}_{ij}) +
\delta({\pmb{r}}_{ij})\,\rho({\pmb{r}})^\beta\, p_{ij}^2 \right\} \nonumber\\
&+&t_5(1+x_5 P_\sigma)\frac{1}{\hbar^2}{\pmb{p}}_{ij}.
\rho({\pmb{r}})^\gamma\,\delta({\pmb{r}}_{ij})\, {\pmb{p}}_{ij}
 \quad ,
\eeqy
where the $t_4$ and $t_5$ terms are density-dependent generalizations of the 
$t_1$ and $t_2$ terms, respectively. The full formalism for this generalized
Skyrme force is presented in the Appendix of Ref.~\cite{cgp09}. With the new 
terms it was possible in
the HFB-18 model to stop the unphysical transition of NeuM (and N*M) to a 
spin-ordered state, with only an insignificant deterioration in the 
quality of the mass fit. Furthermore, in the HFB-18 model our adjustment
of these new terms was minimal, in the sense that the three parameters of the 
$t_5$ term were chosen in a rather arbitrary way, subject only to the condition
of stopping the unwanted transition, while the three parameters of the $t_4$
term were fixed by the equations 
\bmlet 
\beqy\label{3a}
\beta = \gamma  \quad ,
\eeqy
\beqy\label{3b}
t_4 = -\frac{1}{3}t_5(5 + 4x_5)  \quad ,
\eeqy
\beqy\label{3c}
x_4 = -\frac{4 + 5x_5}{5 + 4x_5} \quad ,
\eeqy
\emlet
which ensure that the $t_4$ term cancels exactly the $t_5$ term in 
{\it unpolarized} homogeneous nuclear matter of any degree of charge asymmetry 
at all densities. This latter condition is not strictly necessary, since it can
be violated significantly without compromising the quality of the mass fit or 
the stability of NeuM and N*M against a transition to a polarized state;    
we imposed it just to simplify the parameter search in the mass fit. Thus the 
prescription that we adopted for determining both the $t_4$ and $t_5$ terms in
model HFB-18 leaves us with considerable flexibility for realizing further 
improvements, and in this article we begin to exploit these possibilities. 

The main defect of the HFB-18 model that we address in this article is its lack 
of flexibility in fitting the EOS of NeuM at zero temperature. In 
developing this model we had no trouble in constraining to the EOS of Friedman 
and Pandharipande~\cite{fp81} (FP), which is based on a variational calculation
using the realistic Urbana $v_{14}$ nucleon-nucleon force with the three-body 
force TNI. However, it is by no means clear that the FP EOS for NeuM is the 
correct one, since some quite different ones that are at least equally 
plausible have been published. For example, the EOS of Akmal
{\it et al.}~\cite{apr98} labeled ``A18 + $\delta\,v$ + UIX$^*$", which
we refer to as APR, is considerably stiffer. This EOS, like FP, is based on a 
variational calculation but uses the realistic
Argonne A18 two-body force and the semi-phenomenological UIX$^*$ three-body
force (see Ref.~\cite{apr98} for details of these forces); there is also a
so-called relativistic boost correction $\delta\,v$.
Actually, while the latter EOS is more recent and more complete, 
there have been some recent experimental indications that it might be too 
stiff~\cite{xiao09}. On the other hand, several realistic EOSs for NeuM that 
are still stiffer than APR have been published. One such striking case is based
on the same A18 two-body force as is APR, but uses a much more realistic 
three-body force and is calculated with the Brueckner-Hartree-Fock (BHF) 
method. This is the EOS labeled ``V18" in Ref.~\cite{ls08}, which we will refer
to as LS2 (a very similar EOS, based on the quantum Monte Carlo method, 
has been published in Ref.~\cite{gan09}). In 
this article we make no assumption as to which of these three EOSs, 
FP, APR or LS2, is closer to reality, but rather generate a family
of three mass models with effective forces constrained respectively to these
three realistic EOSs. In this way we may reasonably hope with our family of
effective forces to cover the range of possibilities left open by the present
limitations of our knowledge. However, we find that as long as the conditions
(\ref{3a}) -- (\ref{3c}) are maintained it is impossible to constrain to any
EOS stiffer than that of FP while maintaining a high quality mass fit. 
We shall see here that by releasing these conditions and exploiting the
degrees of freedom associated with the $t_4$ and $t_5$ terms we can stiffen the
EOS of NeuM without sacrificing any of the achievements of the HFB-18 model. 

Other problems that appear to be inevitable in mass models based on the 
conventional form of Skyrme forces include an incorrect distribution of the 
total potential energy among the four two-body spin-isospin channels in 
symmetric nuclear matter (SNM), and Landau parameters with values indicating 
various unphysical instabilities in homogeneous nuclear matter. The new terms
in $t_4$ and $t_5$ allow us enough flexibility to handle both of these problems
as well as that of the stiffness of the EOS of NeuM.

Accordingly, we present in this article three new mass models, HFB-19, HFB-20, 
and HFB-21. The first of these, HFB-19 has its underlying force, BSk19, 
constrained, like BSk18 (the force of model HFB-18), to the FP neutron-matter 
EOS, while BSk20 and BSk21 (the forces of models HFB-20 and HFB-21, 
respectively), are constrained to the APR and LS2 neutron-matter EOSs, 
respectively. All three satisfy the required stability properties of nuclear 
matter, and a qualitatively acceptable distribution of the potential energy 
among the spin-isospin channels. In Sec~\ref{model} we describe the new 
models and many of their properties, with discussions of their symmetry
energy and Landau parameters being treated separately in Secs~\ref{symm} 
and~\ref{llandau}, respectively. The behavior of these models in N*M is 
dealt with in Sec~\ref{nsm}. We summarize our conclusions and discuss 
possible future developments in Sec~\ref{concl}. Appendix~\ref{gognycoll}
describes some microscopic calculations that support our more phenomenological
collective corrections, 
while Appendix~\ref{landau} gives the expressions for the Landau parameters
of neutron matter with our generalized Skyrme forces. 

\section{The HFB-19, HFB-20, and HFB-21 mass models}
\label{model}

\subsection{Principal features of models}
\label{2A}

Our new HFB calculations for finite nuclei are performed almost 
exactly as for the HFB-18 model \cite{cgp09}. In particular, the treatment of 
pairing is highly realistic. As usual, we take a contact 
pairing force that acts only between nucleons of the same charge state $q$, 
\beqy
\label{4}
v^{\rm pair}_q(\pmb{r_i}, \pmb{r_j})= 
v^{\pi\,q}[\rho_n(\pmb{r}),\rho_p(\pmb{r})]~\delta(\pmb{r}_{ij})\quad ,
\eeqy
where the strength $v^{\pi\,q}[\rho_n,\rho_p]$ is a functional of both the 
neutron and proton densities. But instead of postulating a simple functional 
form for the density dependence, as is usually done, we construct the pairing 
force by solving the HFB equations in uniform asymmetric nuclear matter with
the appropriate neutron and proton densities, requiring that the resulting gap 
reproduce exactly, as a function of density, the microscopic $^1S_0$ pairing 
gap calculated with realistic forces~\cite{cao06}. We follow our usual practice
in finite nuclei of allowing the proton pairing strength to 
differ from the neutron pairing strength, and for allowing each of these
strengths to depend on whether there is an even or odd number of nucleons of
the charge type in question. These extra degrees of freedom are taken into
account by multiplying the value of $v^{\pi\,q}[\rho_n, \rho_p]$, as
determined by the nuclear-matter calculations that we have just described,
with renormalizing factors $f^{\pm}_q$, where $f^+_p, f^-_p$ and $f^-_n$ are 
free, density-independent parameters to be included in the mass fit, and  
we set $f^+_n = 1$. (For more details see Refs.~\cite{cgp08,gcp09}.) 

To the HFB energy calculated for the Skyrme and pairing forces we add 
a Wigner correction, 
\beqy\label{5}
E_W = V_W\exp\Bigg\{-\lambda\Bigg(\frac{N-Z}{A}\Bigg)^2\Bigg\}
+V_W^{\prime}|N-Z|\exp\Bigg\{-\Bigg(\frac{A}{A_0}\Bigg)^2\Bigg\} \quad ,
\eeqy
which contributes significantly only for light nuclei ($A < A_0$) or nuclei
with $N$ close to $Z$. 
Our treatment of this correction is purely phenomenological, although physical 
interpretations of each of the two terms can be made \cite{sg02,cgp08}.

A second correction that must be made is to subtract from the HFB energy
an estimate for the spurious collective energy. As described in 
Ref.~\cite{cgp08}, the form we adopt here is
\beqy\label{6}
E_{coll}= E_{rot}^{crank}\Big\{b~\tanh(c|\beta_2|) +
d|\beta_2|~\exp\{-l(|\beta_2| - \beta_2^0)^2\}\Big\} \quad  ,
\eeqy
in which $E_{rot}^{crank}$ denotes the cranking-model value of the rotational
correction and $\beta_2$ the quadrupole deformation, while all other parameters
are fitted freely. While the first term here represents the rotational
correction, phenomenologically modified from its cranking-model value, the
second term takes account of the {\it deformation dependence} of the 
vibrational correction [since $E_{coll}$ as given by Eq.~(\ref{6}) vanishes for 
spherical nuclei we must suppose that the vibrational correction for such 
nuclei is absorbed into the fitted force parameters]. We have 
tested~\cite{sg05} the reliability of the rotational part of this expression 
against an exact calculation made by Bender {\it et al.}~\cite{bend04}, while 
in Ref.~\cite{sg07} 
we discussed the validity of our vibrational correction in terms of the 
constraints provided by shape isomers and fission barriers. Further validation
of our collective correction comes from the microscopic calculations described 
in Appendix~\ref{gognycoll}.

The final correction that we make is to drop Coulomb exchange. This is a device
that we have successfully adopted in our most recent models,
beginning with HFB-15 \cite{gp08},  and it can be interpreted as
simulating neglected effects such as Coulomb correlations, charge-symmetry 
breaking of the nuclear forces, and vacuum polarization. 

The only difference between the present HFB calculations and those for the
HFB-18 model \cite{cgp09} is that we now drop all the terms in $J^2$ and 
$J_q^2$ from the Hamiltonian density (A3) of Ref.~\cite{cgp09}, as is done in 
most parametrizations of the Skyrme force. We were able to accommodate these 
terms in our HFB codes without any problem, and our initial 
motivation for dropping them here was simply to make the forces more compatible
with our EOS code for the inner crust of neutron stars \cite{onsi08}, where 
the inclusion of these terms would have led to considerable complexity.
However, omitting the time-even $J^2$ and $J_q^2$ terms requires, on the 
grounds of gauge invariance~\cite{doba95}, that we drop also the time-odd terms
in $({\bf s_n} + {\bf s_p})\cdot({\bf T_n} + {\bf T_p})$ and $({\bf s_n} - 
{\bf s_p})\cdot({\bf T_n} - {\bf T_p})$. This is ensured in the notation of
Ref.~\cite{bend02} by setting $C_0^T = C_1^T= 0$. In dropping 
these terms from the mass fit, which arise jointly from the $t_1, t_2, t_4$ and
$t_5$ terms, their effect is to some extent absorbed by all the Skyrme 
parameters, notably $W_0$ (compare the values of this parameter shown in
Table \ref{tab1} for force BSk18 with those shown for BSk19 -- BSk21). 
In this way the contribution of the omitted terms will be at least partially 
simulated in the EOS calculations. At the same time, it will be seen in the 
following that this approximation does not lead to any deterioration
in the global fit to masses or radii, although there are implications
for properties depending on the time-odd terms in the Hamiltonian. In the 
present article such properties include the EOS of polarized nuclear matter,
the distribution of the potential energy among the different spin-isospin 
channels, and the Landau parameters. The time-odd terms also play a role in the
exact treatment of the masses of odd nuclei, but not in the equal-filling 
approximation~\cite{pmr08}, which we adopt here, as in all our previous articles. 

\subsection{The data fits}

The parameters of the three new models, i.e., of the Skyrme and pairing
forces, and of the Wigner and collective corrections, are fitted to the same 
set of mass data as was the HFB-18 model, i.e., the 2149 measured masses of 
nuclei with $N$ and $Z \ge$ 8 given in Ref.~\cite{audi03}. In 
making these fits we imposed the NeuM constraints discussed in Sec I, with 
the FP EOS~\cite{fp81} defining force BSk19, the APR EOS~\cite{apr98} defining 
BSk20, and the LS2 EOS~\cite{ls08} defining BSk21 (throughout this article we 
assume zero temperature). The fits were also subject to our usual requirement 
that the isoscalar effective mass $M_s^*$ take the realistic value of 0.8$M$ in 
SNM at the equilibrium density $\rho_0$ (see the 
discussion in Ref.~\cite{sg03}); note that the isovector effective mass $M_v^*$
has no role to play in SNM, as can be seen from Eq.~(\ref{10})). We likewise
imposed a value of $J$ = 30 MeV on the symmetry coefficient; this choice is
certainly consistent with all the available evidence coming from both 
experiment and theory, although the situation is somewhat ambiguous (see 
especially Sec~\ref{JL}). Also we required that the incompressibility $K_v$ 
fall in the experimental range 240 $\pm$ 10 MeV~\cite{col04}. The 
values of the Skyrme, pairing and Wigner parameters resulting from these three
fits are shown in Table~\ref{tab1} ($\varepsilon_{\Lambda}$ is the pairing
cutoff parameter \cite{cgp08,gcp09}). The parameters of the collective 
correction of Eq.~(\ref{6}) are shown in Table~\ref{tab2}. (All the expressions
corresponding to the generalized Skyrme force (\ref{2}) required here are given
in, or can be trivially derived from, the Appendix of Ref.~\cite{cgp09}.) 

\begin{figure}
\centerline{\epsfig{figure=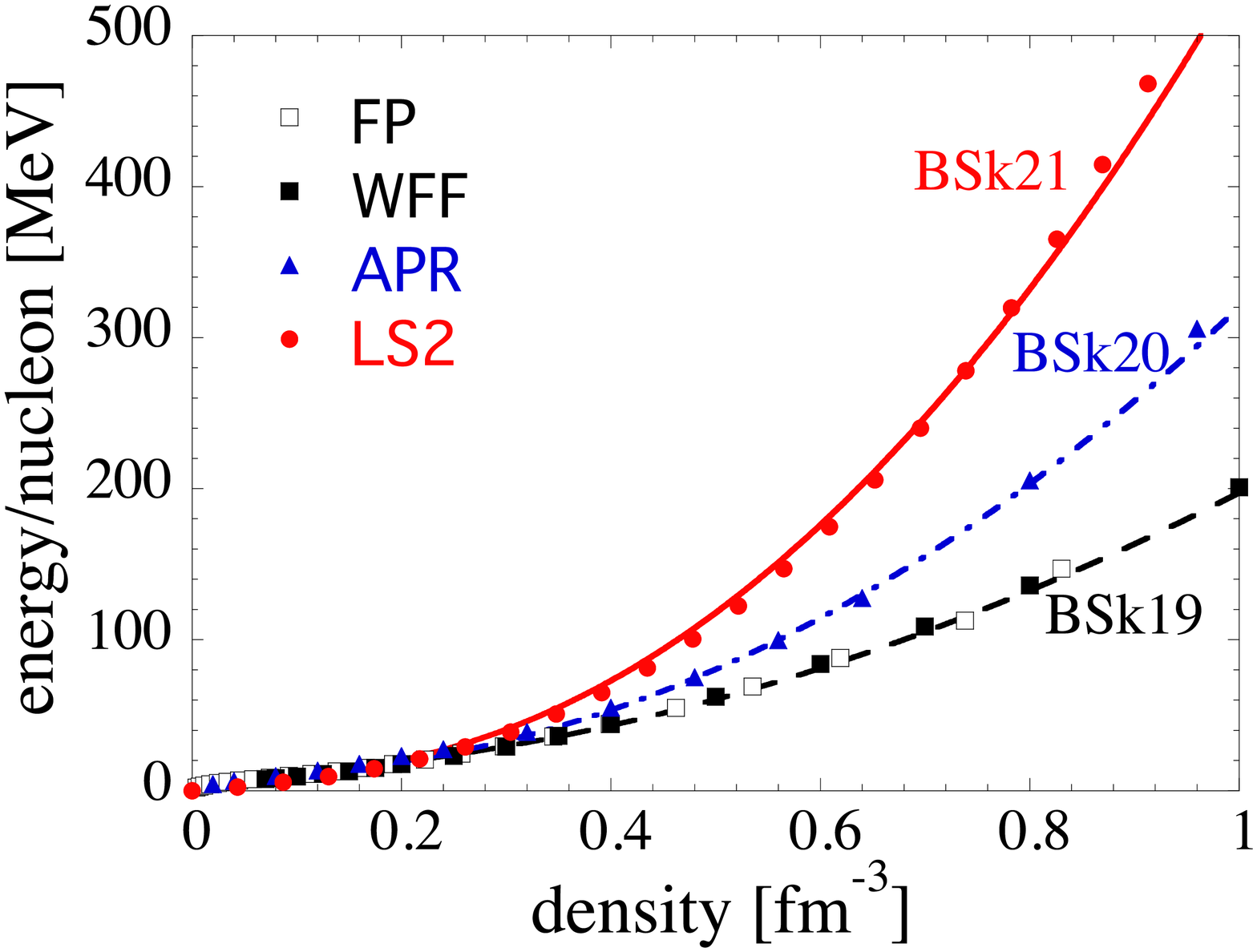,height=8.0cm}}
\caption{(Color online) Zero-temperature EOSs for neutron matter (NeuM) with 
forces BSk19, BSk20 and BSk21. Also shown are the realistic EOSs 
FP~\cite{fp81}, WFF~\cite{wir88}, APR~\cite{apr98} and LS2~\cite{ls08}.} 
\label{fig1}
\end{figure}

Figure~\ref{fig1} shows that each of the three new forces of this article,
BSk19 -- BSk21, reproduces very well its realistic ``target'' EOS of NeuM. In 
this figure we show also the results of the realistic calculation 
``UV14 plus TNI" of Wiringa {\it et al.}~\cite{wir88}, labeled here as WFF. 
This EOS uses the same realistic forces as FP, and like FP is based on a
variational calculation; it agrees very closely with the EOS of FP, 
but runs to higher 
densities, and so generally we shall henceforth take WFF rather than FP as the 
reference EOS for BSk19. We have checked for all of our forces that the ground 
state of NeuM is unpolarized over the entire density range shown in 
Figure~\ref{fig1}, that is, there is no ferromagnetic instability. At the same 
time, in fitting the force parameters we have also to check that the ground 
state of N*M is likewise stable against polarization, since the stability of 
NeuM does not guarantee that of N*M (see Sec \ref{nsm} for further details 
of the N*M calculations). It is seen from Table~\ref{tab1} that
we have relaxed the conditions (\ref{3a}) - (\ref{3c}) that we imposed on the 
search for model HFB-18 \cite{cgp09}, which means that the $t_4$ and $t_5$ 
terms no longer exactly cancel in unpolarized homogeneous nuclear matter of 
arbitrary charge symmetry.  This is the source of the extra flexibility that we
have gained in the present work, although it makes the fitting process much 
more laborious. 

\begin{figure}
\centerline{\epsfig{figure=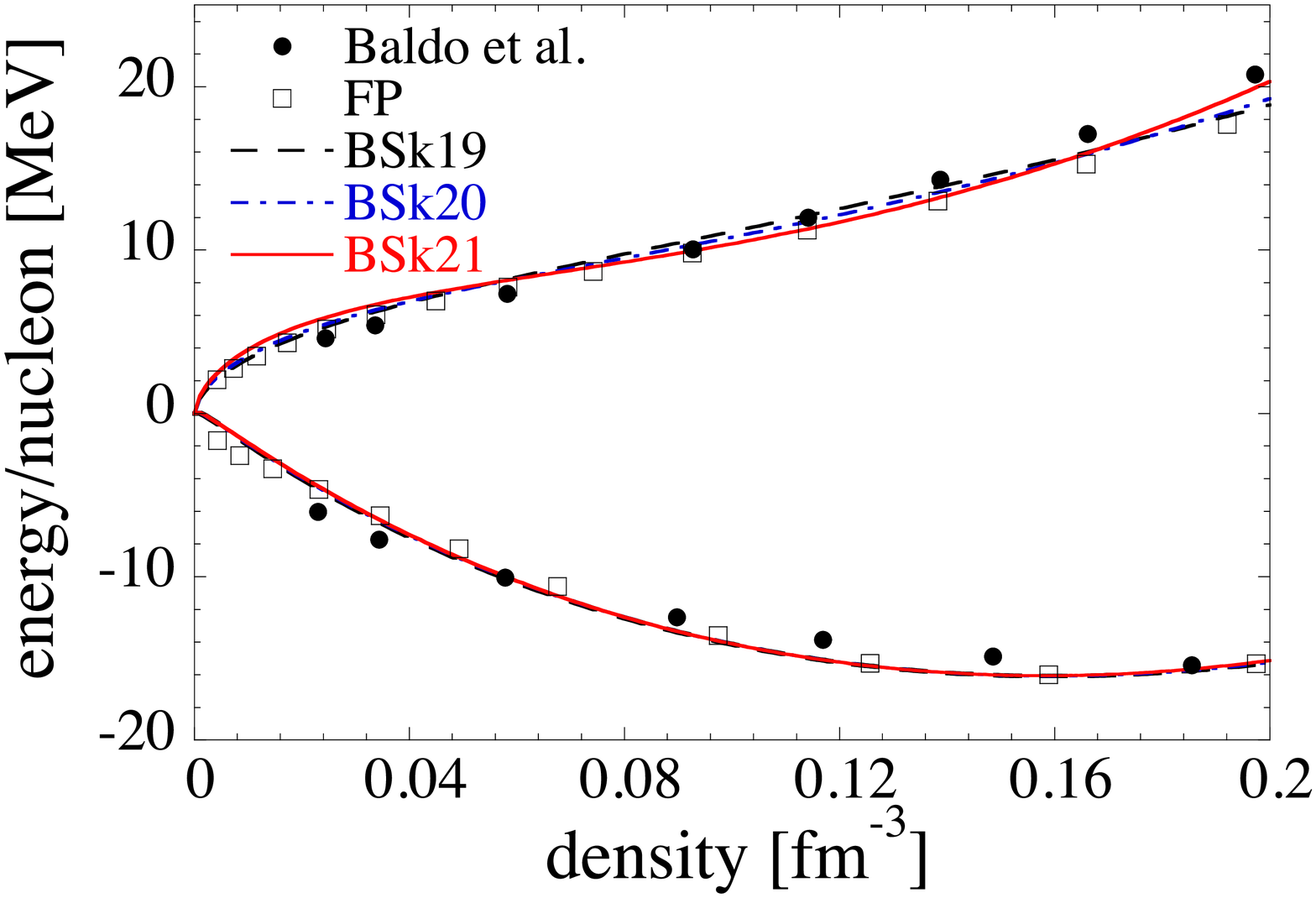,height=8.0cm}}
\caption{Low-density EOS in NeuM and SNM of forces BSk19 -- BSk21. We compare
with the calculations of Refs.~\cite{fp81,bald08}.}
\label{lowrho}
\end{figure}

Figure~\ref{lowrho} is a low-density zoom of Figure~\ref{fig1}, comparing our
forces with the NeuM of FP and also that given by the BHF calculations of 
Baldo~{\it et al.}~\cite{bald08}
(the calculations of WFF are not taken to such low densities). This figure
also shows the corresponding results for SNM. The excellent agreement of all
our forces with the realistic calculations, which have served as the basis of
a density functional for nuclear-structure calculations~\cite{bald08}, 
will be seen. 

The rms and mean (data $-$ theory) values of the deviations between the measured
masses and the predictions for the HFB-19, HFB-20 and HFB-21 models are given 
in the first and second lines, respectively, of Table~\ref{tab3}, where we also
compare with HFB-18~\cite{cgp09}. With HFB-21 we have achieved our best mass 
fit ever,
although we do not regard its superiority over the other three forces as 
significant. The next two lines of Table~\ref{tab3} show the deviations for the
subset consisting of the most neutron-rich measured nuclei, here taken as those
with a neutron separation energy $S_n \le $ 5.0 MeV. All four models display a 
slight deterioration as we move into the neutron-rich region. (The rms 
deviation given by the finite-range droplet model (FRDM)~\cite{frdm} for this 
same subset is 0.910 MeV.) Lines 5 -- 8 show the deviations for the 
$S_n$ and $\beta$-decay energies $Q_\beta$ of all measured nuclei; these 
differential quantities are of greater astrophysical relevance than the 
absolute masses, for both the $r$-process and the inner crust of neutron stars. 
It will be seen that all models fit the $S_n$ better than they fit the 
absolute masses. However, this is the case for the $Q_\beta$ only with model 
HFB-18. Overall, all four models give mass fits of very similar quality. 
Likewise, from lines 9 and 10 we see that all four models are essentially 
equivalent from the standpoint of charge radii (the data are taken from 
Ref.~\cite{ang04}). The last line of Table~\ref{tab3} shows the calculated 
values of the neutron-skin thickness of $^{208}$Pb; the experimental values
are discussed in Sec~\ref{JL}.
 
\begin{figure}
\centerline{\epsfig{figure=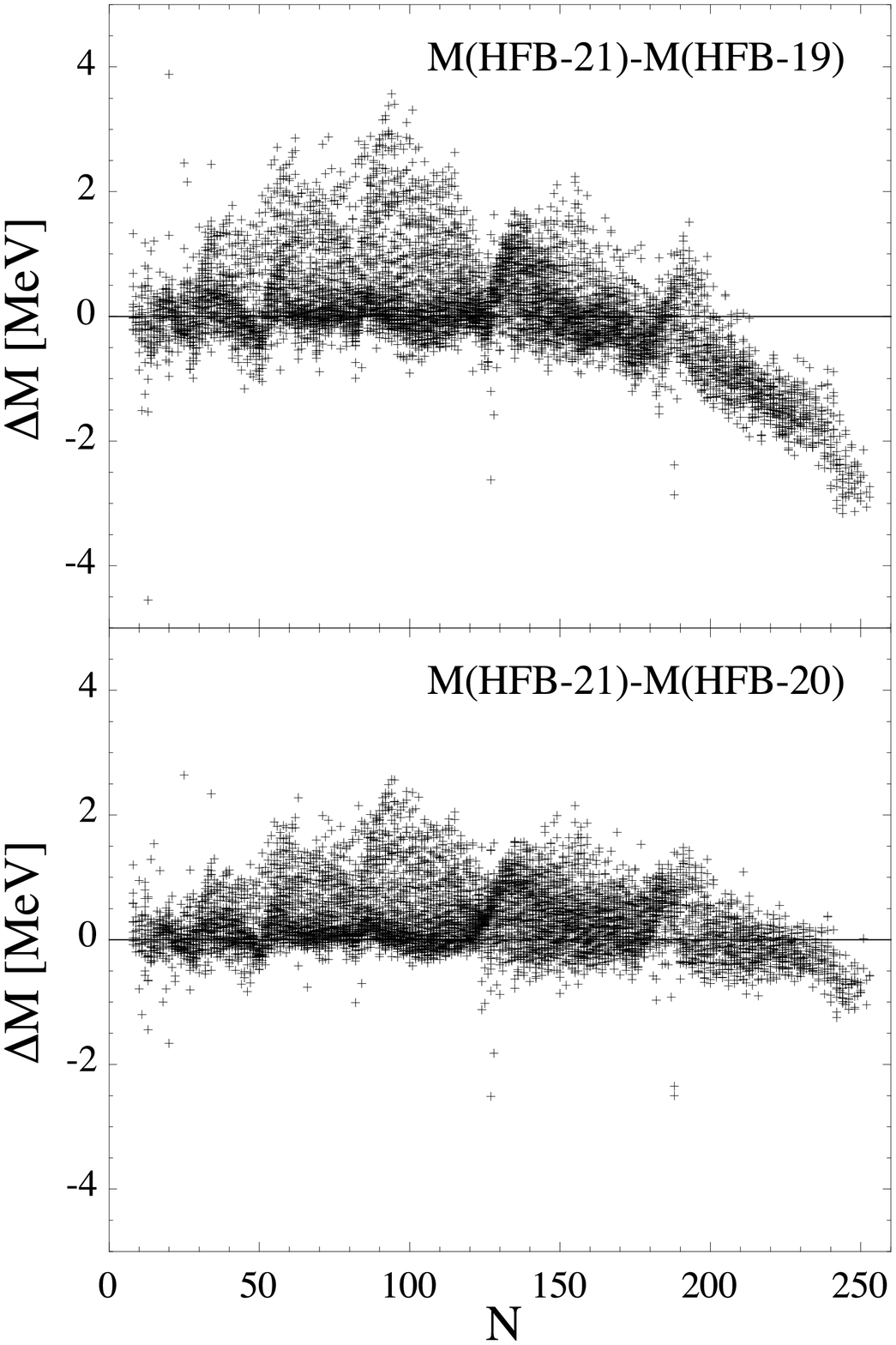,height=8.0cm}}
\caption{(Upper panel) Mass differences between HFB-21 and HFB-19 mass models
for all 8509 nuclei included in the tables. (Lower panel) Ditto for
HFB-21 and HFB-20 models.}
\label{mdif1}
\end{figure}

\begin{figure}
\centerline{\epsfig{figure=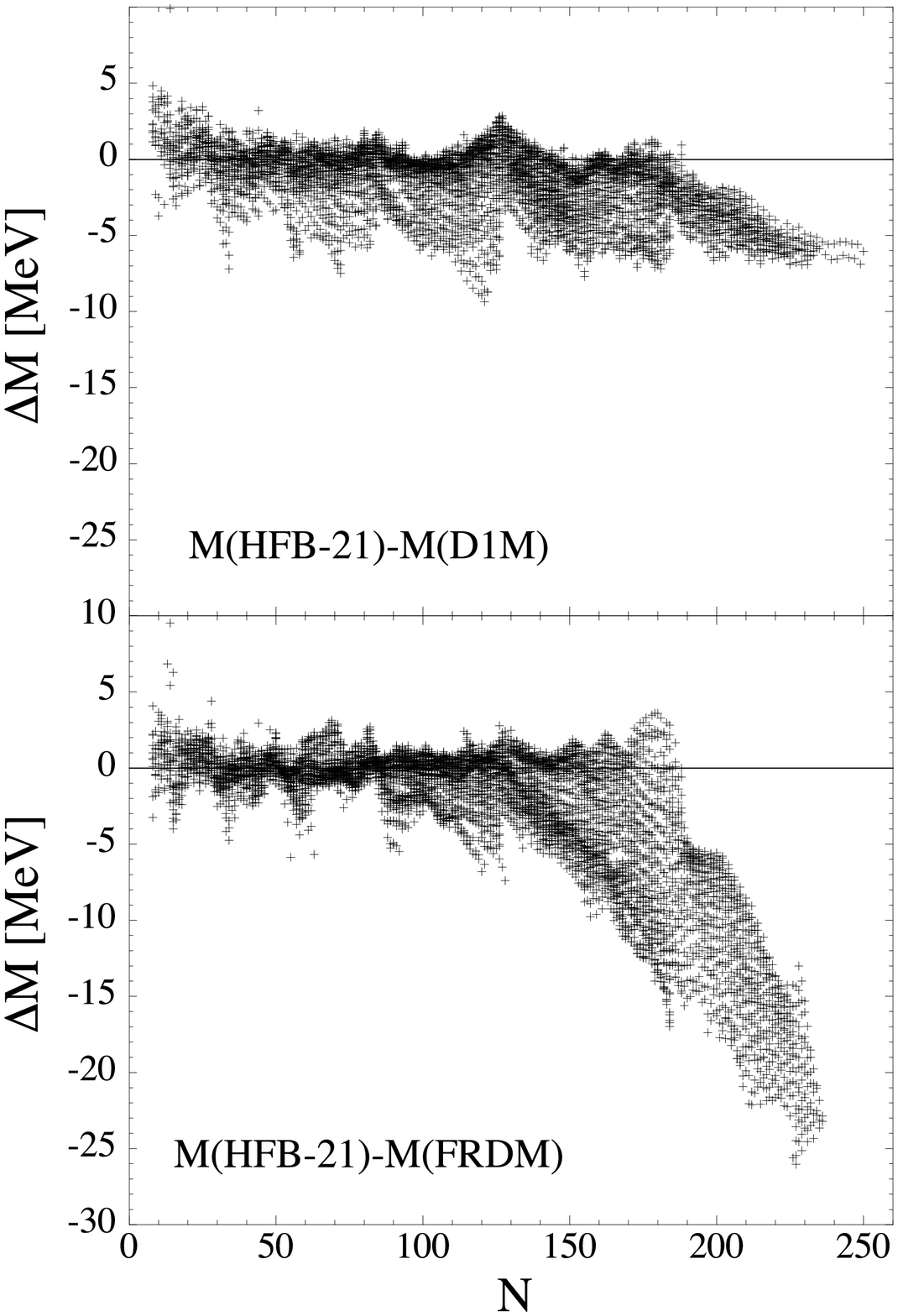,height=8.0cm}}
\caption{(Upper panel) Mass differences between HFB-21 and D1M mass models for 
all 8509 nuclei included in the tables. (Lower panel) Ditto for
HFB-21 and FRDM models.}
\label{mdif2}
\end{figure}

Using the forces BSk19, BSk20 and BSk21 we have constructed complete mass 
tables HFB-19, HFB-20 and HFB-21, respectively, running from one drip line to 
the other over the range $Z$ and $N \ge$ 8 and $Z \le$ 110. We plot the 
differences between these three mass models in Figure~\ref{mdif1}, where it 
will be seen that globally they are very similar, with no striking systematic 
differences emerging even for the highest values of $N$, corresponding to the 
neutron drip line. On the other hand, the upper panel of Figure~\ref{mdif2} 
shows that there are somewhat larger deviations between our models 
(represented here by the typical HFB-21) and the model D1M~\cite{gg09} based on
a Gogny-type force (note the different scale used in Figure~\ref{mdif2}); 
moreover, with increasing $N$ there is a systematic tendency for D1M to bind 
less strongly than the HFB models (the rms deviation of this model for the same 
data set that we take here is 0.798 MeV). The lower panel of Figure~\ref{mdif2} 
compares HFB-21 with the FRDM~\cite{frdm}, and here we see 
that at large $N$ the deviations are much bigger than is the case for D1M. 

\subsection{Properties of infinite nuclear matter}

The first seven parameters of Tab.~\ref{tab4} are defined by first writing 
the energy per nucleon of infinite nuclear matter of density $\rho$ and 
charge asymmetry $\eta = (\rho_n - \rho_p)/\rho$ in the form  
\beqy
\label{7}
e(\rho,\eta)=e(\rho,\eta=0)+ e^{(1)}_{sym}(\rho)\eta^2+ O\Big(\eta^4\Big) 
\quad ,
\eeqy
in which the first term on the right-hand side is just the energy per nucleon
of SNM; we have neglected charge-symmetry breaking terms, such as those 
arising from the neutron-proton mass difference (note that in  
Ref.~\cite{cgp09} we wrote $e^{(1)}_{sym}$ simply as $e_{sym}$, but here
we have to distinguish it from another symmetry energy to be defined in 
Eq.~(\ref{3.1})). We then expand $e(\rho,\eta=0)$ and 
$e^{(1)}_{sym}(\rho)$ about the SNM equilibrium density $\rho_0$ in powers of 
$\epsilon = (\rho - \rho_0)/\rho_0$, thus
\bmlet
\beqy \label{8a}
e(\rho,\eta=0) = a_v + \frac{1}{18}K_v\epsilon^2 - 
\frac{1}{162}K^\prime\,\epsilon^3 + ...
\eeqy
and
\beqy \label{8b}
e^{(1)}_{sym}(\rho) = J + \frac{1}{3}L\epsilon + \frac{1}{18}K_{sym}\epsilon^2
+ ... \quad .
\eeqy 
\emlet
Since the expressions corresponding to the generalized Skyrme force (\ref{2}) 
for the coefficients $K^\prime $ and $K_{sym}$ were not given in the Appendix 
of Ref.~\cite{cgp09} we show them here:
\bmlet
\beqy\label{9a} 
K^\prime = -\frac{12\hbar^2}{5M}k_F^2 + \frac{3}{8}\left[3t_1 +
t_2(5 + 4x_2)\right]\rho\,k_F^2
- \frac{27}{16}(\alpha + 1)\alpha(\alpha - 1)t_3\rho^{\alpha + 1} \nonumber \\
- \frac{9}{80}(3\beta + 5)(3\beta + 2)(3\beta - 1)t_4\rho^{\beta + 1}k_F^2
\nonumber \\
- \frac{3}{80}(3\gamma + 5)(3\gamma + 2)(3\gamma - 1)t_5(5 + 4x_5)
\rho^{\gamma + 1}k_F^2 
\eeqy
and
\beqy\label{9b}
K_{sym} = -\frac{\hbar^2}{3M}k_F^2 + \frac{5}{12}\left[-3t_1x_1 +
t_2(4 + 5x_2)\right]\rho\,k_F^2
- \frac{3}{16}\alpha(\alpha + 1)t_3(1 + 2x_3)\rho^{\alpha + 1} \nonumber \\
- \frac{1}{8}(3\beta + 5)(3\beta+2)t_4x_4\rho^{\beta + 1}k_F^2
+ \frac{1}{24}(3\gamma + 5)(3\gamma+2)t_5(4 + 5x_5)\rho^{\gamma + 1}k_F^2
\quad .
\eeqy
\emlet
The coefficients appearing in lines 8 and 9 of Table~\ref{tab4} are functions
of the preceding coefficients, and are defined in Eqs.~(\ref{3.6}) and
(\ref{3.7}), respectively. They will be used in Sec~\ref{symm}.

{\it Effective masses}. The values of the isoscalar and isovector effective 
masses, $M_s^*$ and $M_v^*$, respectively, shown in Table~\ref{tab4}, are
calculated at the equilibrium density $\rho_0$ of SNM. The value $M_s^* = 0.8M$
for all the models was, as already explained, a constraint imposed on the 
fit to the data, but the various values of $M_v^*$ were left free and emerged 
from the fit. Experimental estimates of this quantity vary widely: 
Measurements of the isovector giant dipole resonance (IVGDR) in heavy nuclei, 
as summarized in
Figure 47 of Ref.~\cite{ber75}, indicate that the value of $M_v^*/M$ can range 
from 0.7 to 1 (this figure in effect plots $M/M_v^*$). 
However, the subsequent discussion in Ref.~\cite{ber75} points out that lower 
values are possible, which means that none of our models is in clear conflict
with experiment. Moreover, all four models have $M_v^* < M_s^*$,
which implies that the neutron effective mass $M^*_n$ is larger than the proton
effective mass $M^*_p$ in neutron-rich matter, since the effective mass of a 
nucleon of charge type $q$ in nuclear matter at density 
$\rho$ is given by 
\beqy\label{10}
\frac{M}{M^*_q} = \frac{2\rho_q}{\rho}\frac{M}{M^*_s} +
\Big(1-\frac{2\rho_q}{\rho}\Big)\frac{M}{M^*_v} \quad .
\eeqy
This prediction  is consistent with measurements of the IVGDR~\cite{les06}, 
and has been confirmed in many-body calculations with realistic 
forces~\cite{lns}. With these latter calculations
giving $M_s^* = 0.825M$ and $M_v^* = 0.727M$, we see that the magnitude of the
splitting given by the new forces, especially BSk21, is much more realistic 
than that given by BSk18. 

\begin{figure}
\centerline{\epsfig{figure=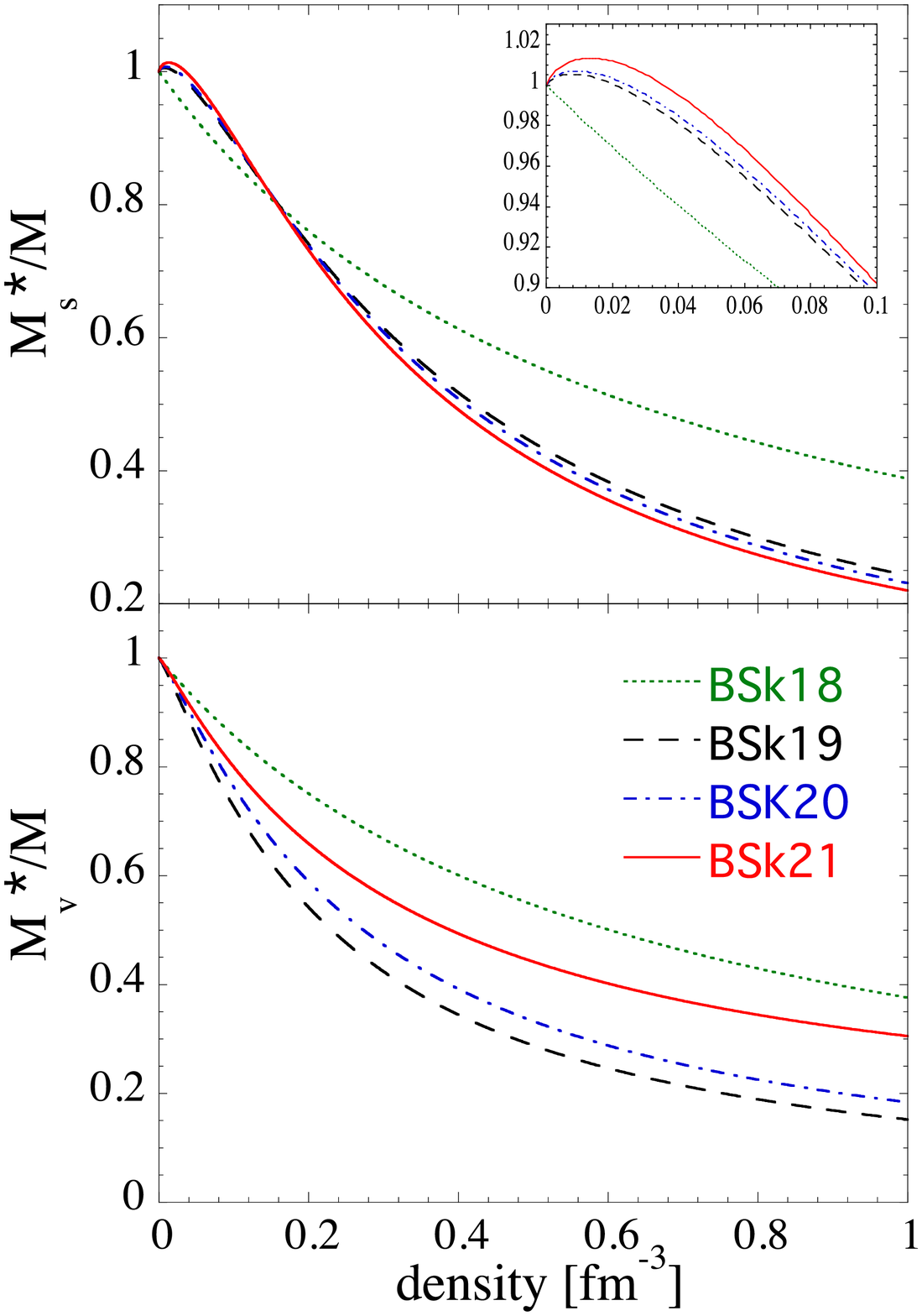,height=8.0cm}}
\caption{(Color online) Variation with density of $M_s^*/M$ (upper panel) and 
$M_v^*/M$ (lower panel).} 
\label{mstar1}
\end{figure}

Figure~\ref{mstar1} shows for the three new models and BSk18 
how $M_s^*$ and $M_v^*$ 
vary with density. For Skyrme forces of the conventional form (\ref{1}) both
$1/M_s^*$ and $1/M_v^*$ depend linearly on the density, as is also the case for 
BSk18, because of the constraints~(\ref{3a}) - (\ref{3c}). However, these 
constraints are not applied to the new forces, BSk19 -- BSk21, with the 
result that $1/M_s^*$ and $1/M_v^*$ will vary non-linearly with density for 
these three forces. This accounts for the low-density peak seen in $M_s^*$ for 
these forces. There are peaks also in $M_v^*$ for all three of the new forces, 
but they occur at such low densities that they are quite invisible in 
Figure~\ref{mstar1}.

\begin{figure}
\centerline{\epsfig{figure=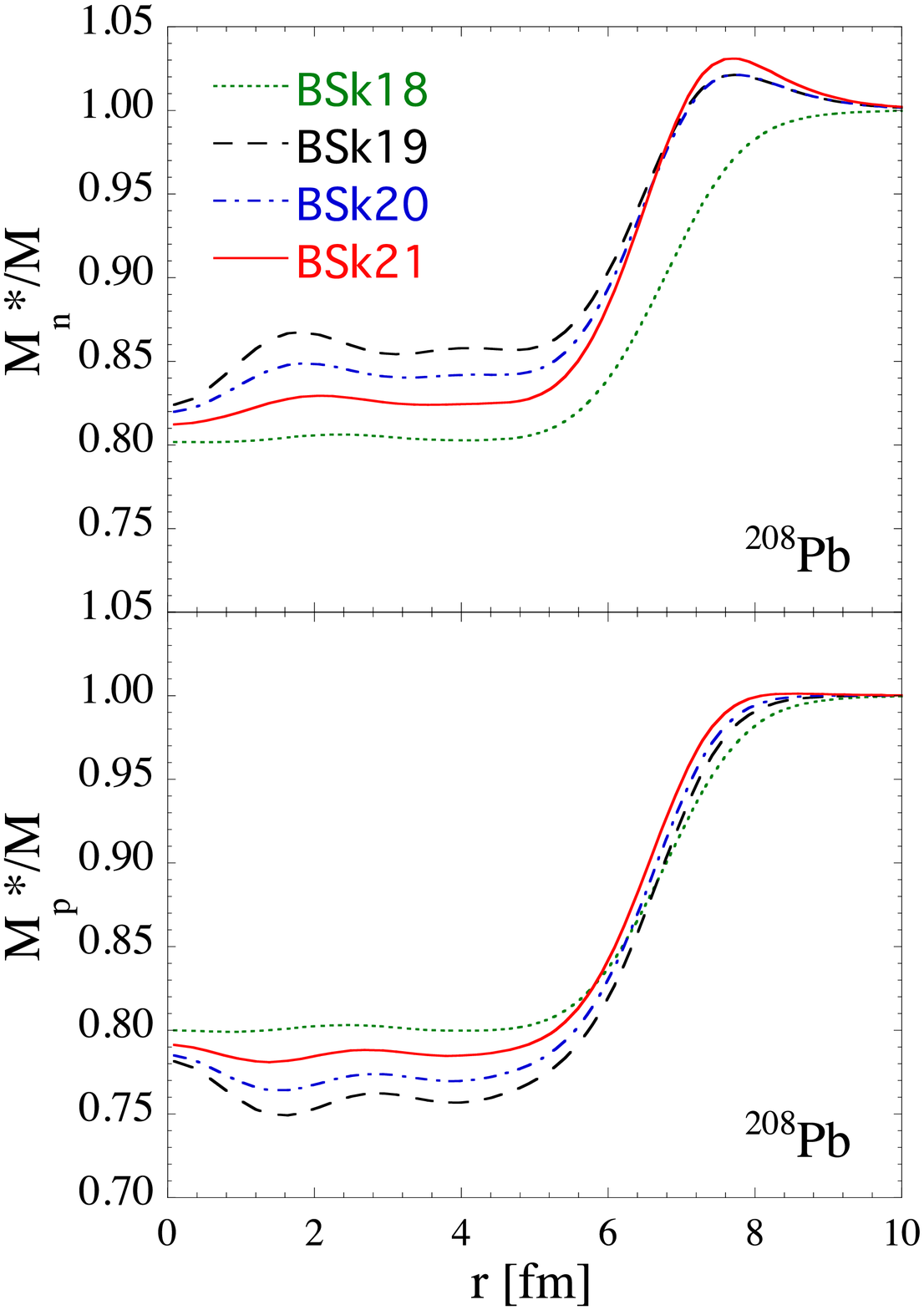,height=8.0cm}}
\caption{(Color online) Variation of $M_n^*/M$ (upper panel) and $M_p^*/M$ 
(lower panel) with radial position in $^{208}$Pb.}
\label{mstar2}
\end{figure}

Using now Eq.~(\ref{10}) and the density distributions given by the HFB 
calculations, we calculate the radial variation of $M_n^*$ and $M_p^*$ 
in $^{208}$Pb, and show the results for the three new models and BSk18 
in Figure~\ref{mstar2}.
We see that the average value of $M_n^*$ over the nucleus of $^{208}$Pb will be
significantly larger for the three new forces than for BSk18, and we would 
therefore expect the neutron single-particle (s.p.) level density close to the 
Fermi surface to be greater for the new forces than for BSk18. 
Table~\ref{tab5} shows that this is indeed the case (of the new forces we show
only BSk20 in this and Table~\ref{tab6} since the other two give almost
identical s.p. energies), if we take as a measure
of the s.p. level density the separation of certain spin-orbit doublets (we 
have to adopt this measure because of the distorting effect of the spin-orbit 
splitting). However, for all four forces the neutron s.p. level density is
still much lower than the experimental value (our data on s.p. energies are
taken from Ref.~\cite{vb72}), essentially because even for the 
new forces the average value of $M_n^*$ over the nucleus is not big enough.
On the other hand, for the proton s.p. level densities the new forces give
just as poor an agreement with experiment as does BSk18 (Table~\ref{tab6}), a 
result that can be understood from the lower panel of Figure~\ref{mstar2}, where 
it will be seen that $M_p^*$ for the new forces lies sometimes above and 
sometimes below the BSk18 value.

Nevertheless, we see that with a more thorough exploitation of the $t_4$ and 
$t_5$ terms it should be possible in principle for the s.p. level densities to 
be well reproduced, at least in heavy nuclei, while maintaining realistic 
values of $M_s^*$ and $M_v^*$ in SNM at density $\rho_0$. Some partial success 
in this direction has 
already been achieved in Ref.~\cite{fpt01}, where the s.p. proton level 
density in the vicinity of the Fermi surface of $^{208}$Pb was well reproduced,
although the s.p. neutron level density was far too low, and no
attempt was made to impose a global mass fit. The peaks in that work were much
stronger than here, $M_s^*$ having a maximum value of 1.13$M$, while $M_v^*$ 
had a maximum value of 1.33$M$, suggesting that forces BSk19 -- BSk21 
do not have nearly enough non-linearity in $1/M_s^*$ and $1/M_v^*$. However, 
sufficiently strengthening the peaks in $M_s^*$ and $M_v^*$ while maintaining
the present quality of the mass fits and respecting all the other constraints
might be impossible with the present form (\ref{2}) of Skyrme force: despite
its great generality it might have to be generalized still further by adding 
extra $t_3, t_4$ and $t_5$ terms, each characterized by different density 
exponents. Fitting such a force would be highly labor intensive.

\begin{figure}
\includegraphics[scale=0.6]{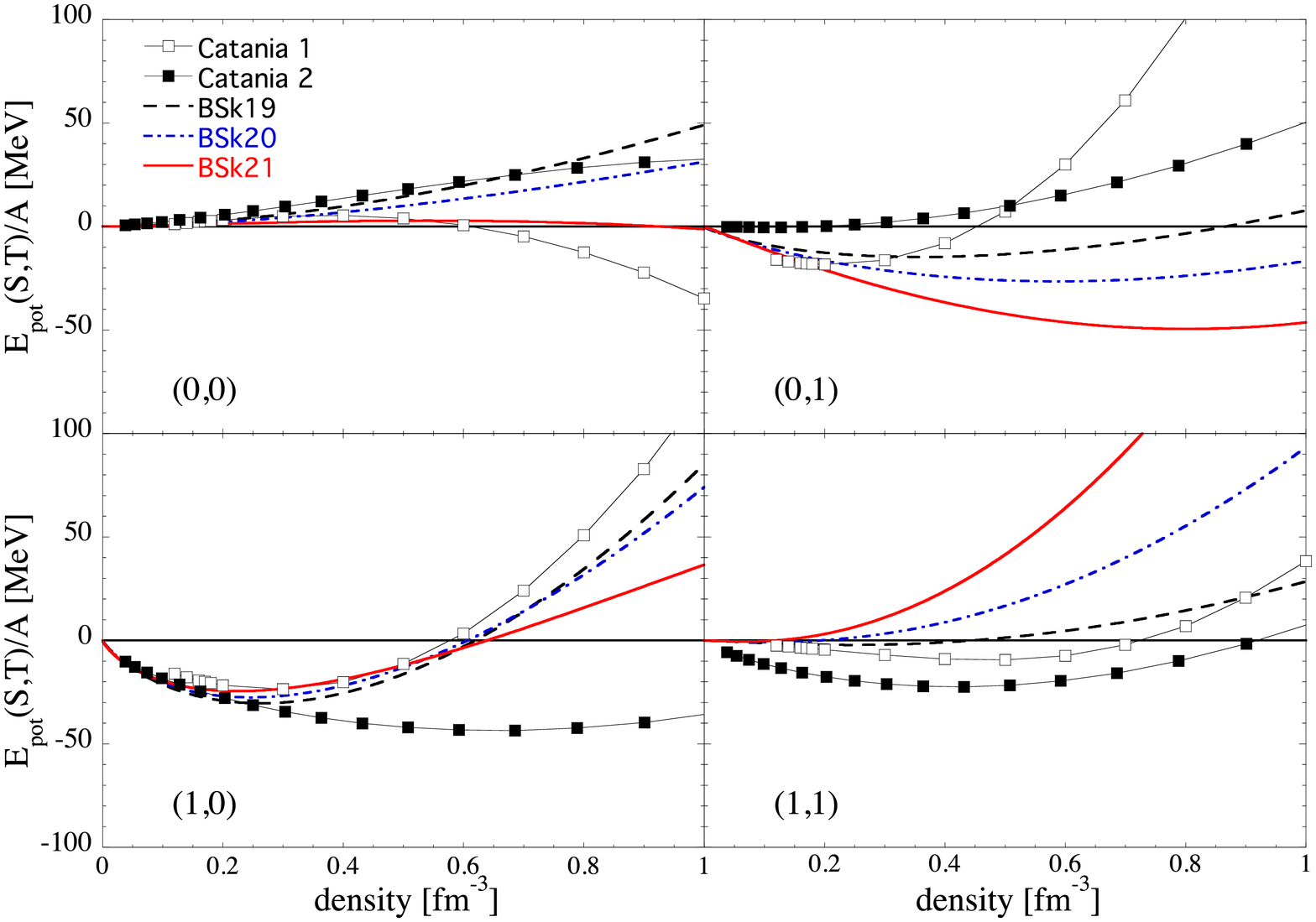}
\caption{(Color online) Potential energy per particle $E_{pot}/A$ in each 
$(S,T)$ channel as a function of
density for symmetric infinite nuclear matter. The open symbols 
correspond to the ``Catania 1" BHF calculations \cite{ls08}, and the solid
symbols to the ``Catania 2" BHF calculations \cite{zhou04}.}
\label{fig_est}
\end{figure}

{\it Distribution of potential energy among the $(S, T)$ channels.} Fitting
our forces to the mass data and the EOS of NeuM is not a sufficient condition 
for ensuring a realistic distribution of the potential energy per nucleon 
among the four two-body spin-isospin $(S, T)$ channels in SNM. 
This is made clear in Figure~\ref{fig_est}, which shows this distribution in SNM
for each of our three new forces, calculated using Eq. (B6) of 
Ref.~\cite{les06}, as a function of density, and compares with two different
BHF calculations: ``Catania 1",  based on Ref.~\cite{ls08} and 
``Catania 2", based on Ref.~\cite{zhou04}. The data of 
the former were kindly supplied by U. Lombardo~\cite{lom08} and of the latter
by M. Baldo~\cite{baldo10}. Given the evident uncertainty in what the real
distribution actually is, the level of agreement we have found with our new 
forces can be regarded as satisfactory. This is the first time that we have 
been able to achieve such a level of conformity to reality with any of our 
forces; the improvement in the (1, 1) channel is particularly striking, as can 
be seen by comparing Figure~\ref{fig_est} with Figure 4 of Ref.~\cite{gcp09} and 
Figure 9 of Ref.~\cite{cgp08}. As discussed in Refs.~\cite{cgp08} 
and~\cite{les06}, this improvement would have been very difficult within the 
framework of conventional Skyrme forces, and the terms in $t_4$ and $t_5$ have 
been indispensable in this respect. 

\section{Symmetry energy}
\label{symm}

\subsection{Generalities}

\begin{figure}
\centerline{\epsfig{figure=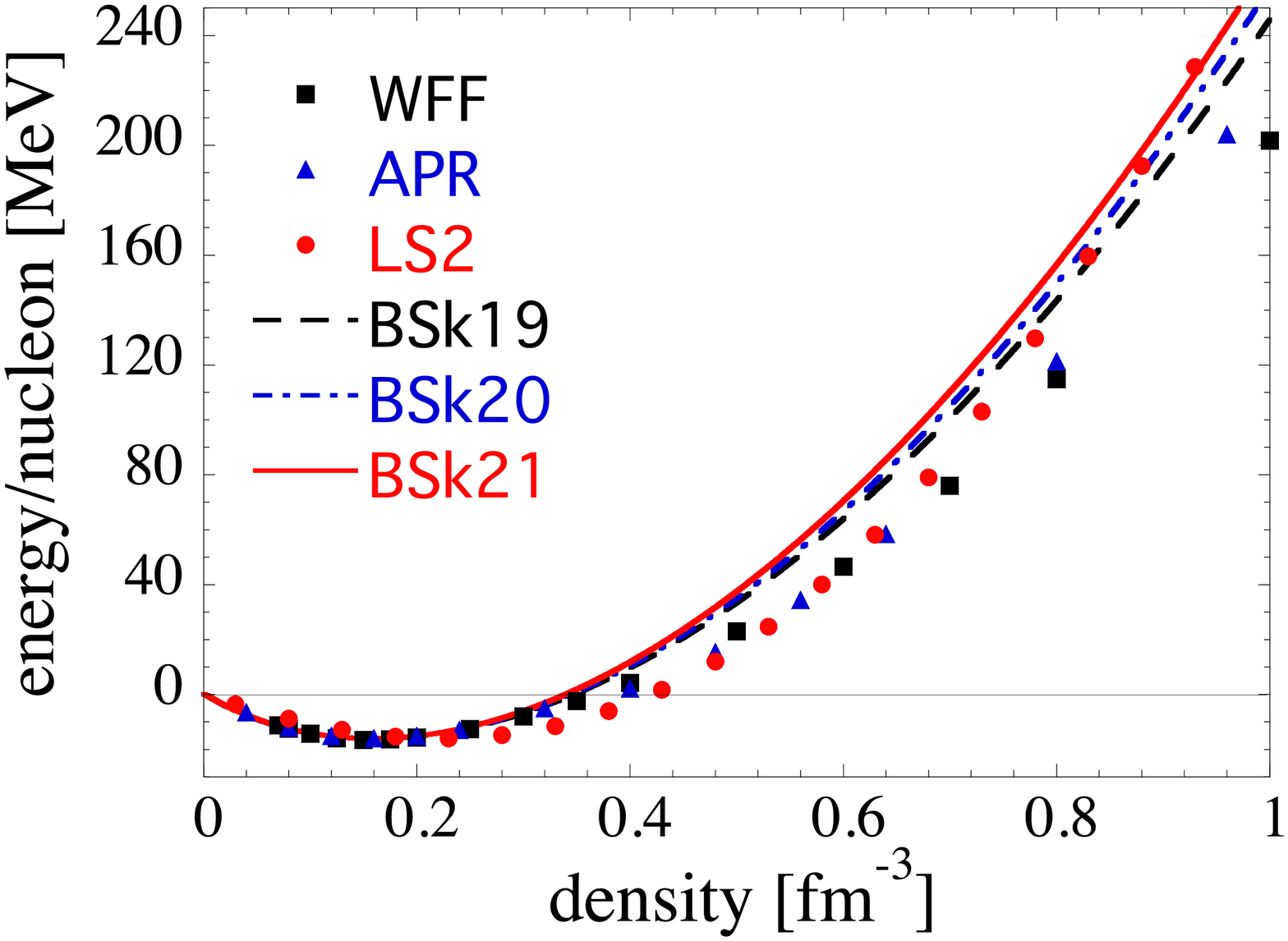,height=8.0cm,angle=0}}
\caption{(Color online) Zero-temperature EOSs for symmetric nuclear matter 
(SNM) with forces BSk19 -- BSk21. Also shown are the realistic EOSs 
WFF~\cite{wir88}, APR~\cite{apr98} and LS2~\cite{ls08}.}
\label{fig3.1}
\end{figure}

The properties of neutron-star matter (N*M) depend crucially on the difference  
between the energy per nucleon in NeuM and the energy per nucleon in SNM, i.e.,
on the symmetry energy
\beqy\label{3.1}
e^{(2)}_{sym}(\rho) = e(\rho,\eta = 1) - e(\rho,\eta = 0)
\eeqy
(see Sec \ref{nsm}). Because of quartic and higher-order terms in $\eta$ 
this symmetry energy is not identical to the symmetry energy 
$e^{(1)}_{sym}(\rho)$ defined in Eq.~(\ref{7}) \cite{ste06}; we return to this 
point later in this section. Figure~\ref{fig3.1} shows that the EOSs in SNM for 
our three new forces, BSk19 -- BSk21, are remarkably similar, despite
having quite different properties in NeuM (see also Figure~\ref{lowrho}). It 
seems that the fit to the data, none of which relates to densities much greater
than $\rho_0$, determines almost completely the properties of SNM up to more 
than 4$\rho_0$, at least under the constraint of given values of $J$ and $K_v$. 
Moreover, we
have calculated the pressure in SNM as a function of density for our forces
and find that it is consistent with measurements of nuclear-matter flow in
heavy-ion collisions, although close to the upper limit: see Figure 3 of 
Ref.~\cite{dan02}.

In Figure~\ref{fig3.1} we show also the realistic EOSs for SNM given by the 
WFF~\cite{wir88}, APR~\cite{apr98} and LS2~\cite{ls08} calculations. It will be 
seen that while BSk19 agrees very closely (by construction) with WFF in
NeuM (and thus with FP~\cite{fp81}), 
it is significantly stiffer in SNM; a similar 
remark applies to BSk20 and its APR counterpart, and likewise to BSk21 and its 
LS2 counterpart. In this connection it should be realized
that the uncertainties in many-body theory are even greater for SNM than for
NeuM, essentially because of the $^3S_1 - ^3D_1$ tensor coupling, which acts
in the former but not the latter. Thus in all such calculations performed so
far it has been necessary to make a phenomenological adjustment of the
three-body force in order to get an acceptable equilibrium point of SNM.
The way in which this adjustment is made is far from unique, both with
regard to the parametrization of the three-body force and the actual
values of the parameters $a_v$ and $\rho_0$ characterizing the equilibrium
point of SNM (we stress that in the case of our forces the values of $a_v$
and $\rho_0$ given in Table~\ref{tab4} were not fixed {\it a priori} but
emerged from the mass fit).

\begin{figure}
\centerline{\epsfig{figure=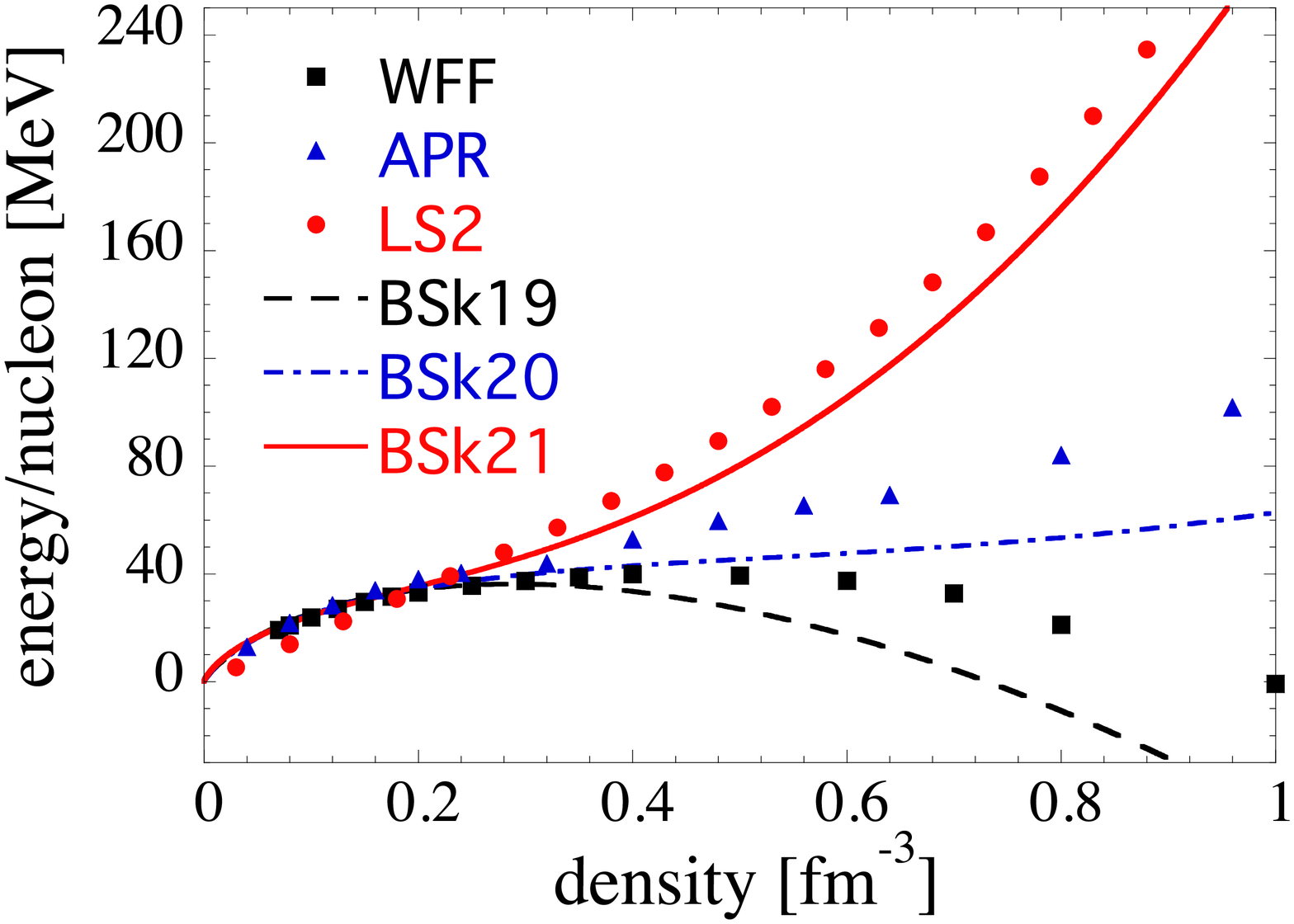,height=8.0cm,angle=0}}
\caption{(Color online) Symmetry energy $e^{(2)}_{sym}$ for forces BSk19, 
BSk20 and BSk21.
Also shown are the realistic EOSs WFF~\cite{wir88}, APR~\cite{apr98} and 
LS2~\cite{ls08}.}
\label{fig3.2}
\end{figure}

Using Figs.~\ref{fig1} and~\ref{fig3.1}, we now plot in Figure~\ref{fig3.2} 
the symmetry energy $e^{(2)}_{sym}$ for all the forces of Figure~\ref{fig3.1}.
We see that our three forces agree very closely in their predictions for 
$e^{(2)}_{sym}$ up to and slightly beyond $\rho_0$.  However, as the density 
increases the curves for our forces diverge strongly from each other. In
fact, for BSk19 $e^{(2)}_{sym}$  becomes negative at high densities, NeuM 
having lower energy per nucleon than SNM. Figure~\ref{fig3.2} also shows that WFF
(and thus FP) has higher symmetry energy than its BSk19 counterpart, and 
likewise for APR as compared to
BSk20, and LS2 as compared to BSk21. These differences can be traced almost 
entirely to the differences in SNM that we have remarked above (note that 
stiffer SNM implies softer symmetry energy, for a given EOS of NeuM).

\begin{figure}
\centerline{\epsfig{figure=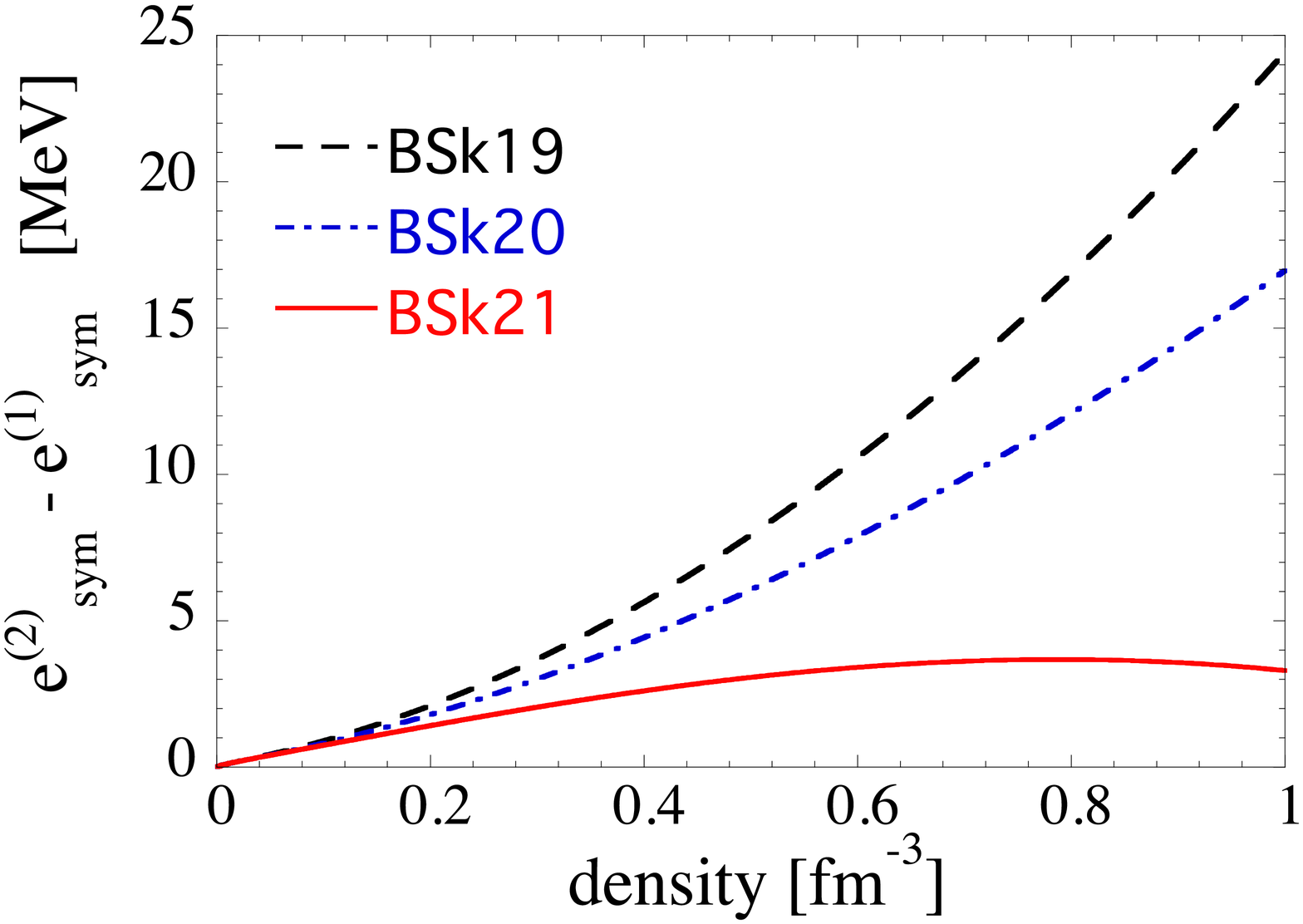,height=8.0cm,angle=0}}
\caption{(Color online) Difference $e^{(2)}_{sym} - e^{(1)}_{sym}$ for 
forces BSk19 -- BSk21.}
\label{fig3.3}
\end{figure}

To examine the relationship between the symmetry energy $e^{(2)}_{sym}$
discussed in this section and the symmetry energy $e^{(1)}_{sym}$ defined in
Eq.~(\ref{7}) we plot in Figure~\ref{fig3.3} the difference 
$e^{(2)}_{sym} - e^{(1)}_{sym}$ for our three new forces. 
Contrary to the usual assumption, this 
difference is seen to be far from negligible, amounting to around 1 MeV in the
vicinity of $\rho = \rho_0$, and possibly becoming much larger at higher 
densities, depending on the force in question.  

{\it High-density behavior.} Since the EOSs of NeuM to which we have fitted our
three new forces differ so strongly at high densities, it would be highly 
desirable to know how the symmetry energy actually does vary with density at 
high density. Unfortunately, as we now see, our knowledge in this respect is 
extremely limited. Turning first to many-body theory with realistic forces, we 
note that all three of our new forces are supported by their realistic 
counterparts. In particular, the plausibility of the negative symmetry energy
of BSk19 is supported by the realistic EOS of WFF. The reason why the realistic
calculations of the symmetry energy show almost as large a high-density 
divergence as our phenomenological forces lies at least partially with the 
choice of three-nucleon force and its behavior at high densities. 
Dirac-Brueckner-Hartree-Fock calculations, in which there is no three-nucleon 
force, also show a high-density softening of the EOS~\cite{ks06}.

In the present state of the theory only experiment can decide the issue, but so
far there has 
been only one measurement of symmetry energy at the high densities found in
the core of neutron stars. This involves pion production in heavy-ion 
collisions at very high energies, measuring the $\pi^-/\pi^+$ 
ratio~\cite{xiao09}. It was found that the data were consistent with models
predicting negative symmetry energy at densities above 3$\rho_0$.

{\it Low-density behavior.} Since $e^{(2)}_{sym}$ is roughly equal to 
$e^{(1)}_{sym}$ for our forces at nuclear and lower densities it can be 
represented by the expansion~(\ref{8b}) in this density region, at least
qualitatively. The close agreement between our 
forces seen in Figure~\ref{fig3.2} for these densities can now be 
related firstly to the fact that $J$ is the same for all our forces (by 
choice), and also to the fact that the $L$ coefficient, which measures the 
slopes of the curves at $\rho = \rho_0$, has similar values, as shown in
Table~\ref{tab4}. Moreover, the origin of the high-density divergence between 
our EOSs can be found in Table~\ref{tab4}, where we see that the second 
derivative of these curves, measured by $K_{sym}$, differs considerably from 
one force to another. 

We devote the rest of this section to our knowledge, both experimental and 
theoretical, of symmetry energy in the region of nuclear densities, since 
much more information is available here than in the high-density region.

\subsection{The $J$ and $L$ coefficients}
\label{JL}

The determination of the $J$ and $L$ coefficients from measurements on finite
nuclei requires the use of some model to describe the nuclei in question. 
One obvious approach is via mass measurements, but fits of mass models to the 
data yield a wide range of values of $J$ and $L$. For example, the 
finite-range droplet model~\cite{frdm} yields $J$ = 32.73 MeV (actually, this
is just for the macroscopic part; in the microscopic part $J$ is given as 35 
MeV). On the other hand, mass models based on Skyrme forces of the 
conventional form~(\ref{1}) give an optimal mass fit with $J$ $\approx$ 
27.5 MeV~\cite{sg05} (we adopt higher values in all our HFB models in order 
to avoid a collapse of NeuM, which would otherwise occur even when assuming a 
non-polarized ground state). However, despite this ambiguity in the value of
$J$ given by mass fits there is a fairly tight correlation with the value of 
$L$. This correlation was explored more than 30 years ago~\cite{fpr78}, and 
then in Ref.~\cite{ton84} (see especially Table 2 of that article); it has 
subsequently been revisited several times, e.g., 
Refs.~\cite{rei99,war09,rei10}.  As for our own HFB mass models, the $L - J$ 
correlation is clearly seen in Table 3 of Ref.~\cite{sg05}, and is manifested 
in the present article by the relative closeness of the $L$ values for our
forces. 

The fact that fitting nuclear masses correlates $J$ with $L$ can be easily 
understood as follows. Increasing $J$ will increase the symmetry energy in 
the center of the nucleus, an increase that will have to be compensated by a 
decrease in the symmetry energy in the surface. Such a decrease can be
achieved by a reduction in $L$, since $\rho < \rho_0$ in that region.

A similar but somewhat weaker $L - J$ correlation is also found in 
measurements of isospin diffusion
in heavy-ion collisions: see Figure 3 of Tsang {\it et al.}~\cite{tsa09}. The
data leave a large range in the possible values of $L$ for a given value of 
$J$, but for our chosen value of $J = 30$ MeV the lower limit on $L$ is 
about 45 MeV, which is consistent with force BSk21, but not the other 
forces. 

We now summarize two types of measurement that lead in principle to unique 
values of either $J$ or $L$. 

{\it Heavy-ion collisions.} Chen {\it et al.}~\cite{ckl05} 
have analyzed isospin-diffusion data and find $L$ = 88 $\pm$ 25 MeV, which is 
much higher than is given by any of our forces. On the other hand, the
analysis of Famiano {\it et al.}~\cite{fam06} finds that the symmetry energy
varies roughly as $\rho^{1/2}$, which yields $L \approx 3J/2$, i.e., 
47 $\pm$ 2MeV, which agrees best with BSk21. The interpretation of these 
experiments is clearly very model dependent.

{\it Neutron-skin thickness.} For a given nucleus this is defined by 
\beqy\label{3.2}
\theta \equiv R_n^{rms} - R_p^{rms} \quad ,
\eeqy
where the rms radii refer to {\it point} nucleons. By considering the results
of non-relativistic Hartree-Fock and relativistic mean-field calculations with
many different forces it was noted by Brown~\cite{bab00} and by Typel and 
Brown~\cite{tb01} that a strong correlation exists between the neutron-skin
thickness and the value of the $L$ coefficient. Thus the strong correlation 
already noted between $L$ and $J$ implies that the value of $\theta$ is 
likewise correlated with $J$: see, for example, Table 2 of Ref.~\cite{ton84}. 
This correlation can easily be understood~\cite{sg05} in terms of the 
droplet-model expression~(2.21) of Ref.~\cite{dm69} for the neutron-skin
thickness of a nucleus of atomic number $Z$ and mass number $A$,
\beqy\label{3.3}
\theta = \frac{3}{2}r_0\frac{J}{Q}I \quad ,
\eeqy
where $r_0 = (3/4\pi\rho_0)^{1/3}$, $I =(N - Z)/A$ and $Q$ is the 
surface-stiffness coefficient, which is anticorrelated with $J$ if masses are
fitted~\cite{fcp80,fcp81}; see also Ref.~\cite{war09} for a recent extensive
discussion. (The correlation between $\theta$ and $J$ might be expected to be
more robust than that between $\theta$ and $L$, since the latter can
hold only to the extent that a local-density approximation is valid in the
surface, $L$ relating to homogeneous nuclear matter.)
    
Measurement of the neutron-skin thickness is difficult, but one particular  
method involves the use of antiprotons as a probe~\cite{jas04}. Analyzing the 
results on 26 nuclei, Ref.~\cite{war09} found $L = 55 \pm 25$
MeV, which is compatible with all three forces of this article. 
 
A nucleus of particular interest is $^{208}$Pb, since several measurements of
its neutron-skin thickness have been made, and a very precise one is being 
planned. It is for this reason that we show in the last line of Table~\ref{tab3}
our calculated values of the neutron-skin thickness for this nucleus. One of 
the lowest published values, obtained from high-energy proton scattering, is 
0.14 $\pm$ 0.04 fm~\cite{hof80}, which certainly agrees with all our forces. A 
similarly low value, 0.15 $\pm$ 0.02 fm, was originally extracted from the 
antiprotonic measurements~\cite{jas04}, but a new analysis of this 
data~\cite{bab07} gives a result that is barely consistent with any of 
our forces: 0.20 $\pm 0.04 \pm 0.05$ fm., in which the first error bar 
represents the experimental uncertainty and the second the theoretical 
uncertainty associated with the strong-interaction model.

This latter source of uncertainty besets, in one form or another, all the
different determinations of neutron-skin thickness published so far, but will
be avoided in the proposed PREX experiment \cite{prex}, which is to measure
parity violation in the elastic scattering of electrons on $^{208}$Pb. 
However, the expected error bars, $\pm$ 0.05 fm., will still be much too 
large to distinguish between our different forces. On the other hand, 
it is conceivable that the PREX experiment will give a result that is quite 
incompatible with any of our predictions. In that case it would be
necessary to make a new mass fit with $J$ constrained to a value appropriately 
higher or lower than the value of 30 MeV taken here. 

{\it Many-body theory with realistic forces.} Several such calculations of 
$J$ and $L$ have been published, but usually these consider only SNM and 
NeuM, and then assume that $e^{(1)}_{sym}$ = $e^{(2)}_{sym}$, an 
approximation that we estimate, on the basis of the forces BSk18-21, to lead
to $J$ being overestimated by around 1 MeV. Further errors in the values of $J$ 
and $L$ calculated in this way arise from the uncertainties that are
specific to the many-body calculation of SNM (see above). Thus it is not
surprising that the values of $J$ and $L$ predicted by different calculations
should disagree widely, as we now see. 

For example, while the three BHF calculations of Ref.~\cite{vid09} have $J$ 
lying in the range 34.7 $\pm 1.1 $ MeV and $L$ in the range 65.0 $\pm 1.9 $ MeV,
the BHF calculations underlying the LS2 EOS (to which BSk21 was constrained) 
yield the quite different values of $J$ = 30.6 MeV and $L$ = 101.4 
MeV~\cite{ls08}. Our final example is the BHF calculation~\cite{ls08} using the 
Bonn B~\cite{mac89} two-nucleon interaction and a three-nucleon interaction of
the same realistic form as adopted for LS2: this calculation gives 
$J$ = 29.4 MeV and $L$ = 74.4 MeV. 

All in all, it is clear that in the search for the correct values of $J$ and 
$L$ many-body theory does not help us to narrow down the wide range of 
possibilities left open by experiment.

\subsection{The $K_{sym}$ coefficient} 

The coefficient $K_{sym}$ appearing in the expansion~(\ref{8b}) can only be 
measured in the combination 
\beqy\label{3.6}
K_\tau = K_{sym} + L\left(\frac{K^\prime}{K_v} - 6\right) \quad,
\eeqy
because for non-zero $L$-values the equilibrium density of
asymmetric nuclear matter is displaced away from $\rho_0$.

{\it Many-body theory with realistic forces.} The only such calculations of
$K_{\tau}$ with which we are familiar are the three BHF calculations of
Ref.~\cite{vid09}, which yield values lying in the range -344 to -335 MeV,
discriminating thereby against BSk21. Of course, the doubts we 
expressed above concerning the reliability of calculations of $J$ and $L$ with 
realistic forces apply equally well to these calculations of $K_{\tau}$.

{\it Isospin diffusion in heavy-ion collisions.} These measurements yield a
value for $K_{\tau}$ of -370 $\pm$ 120 MeV~\cite{chenetal}, which is
consistent with all our forces (see Table~\ref{tab4}), and with the {\it ab
initio} calculations of Ref.~\cite{vid09}. Clearly, improved accuracy will be 
necessary before it is possible to discriminate between our different forces
on this basis.

{\it Isotopic variation of ``breathing-mode" energies.} The coefficient 
$K_{\tau}$ can in principle be determined by measuring the finite-nucleus 
incompressibility $K(Z, A)$ of a string of isotopes of the same 
element. Such a determination was made recently in Ref.~\cite{lietal} by 
measurements of the energies of the giant isoscalar monopole resonance 
$E_{GMR}$ in the seven even-even Sn isotopes between $^{112}$Sn and 
$^{124}$Sn. Using the relation
\beqy\label{3.4}
K(Z, A) = \frac{M}{\hbar^2}R^2E_{GMR}^2  \quad ,
\eeqy
where R is the rms matter radius, they fitted their results 
to the simple expression \cite{bla80}
\beqy\label{3.5}
K(Z, A) = K_v + K_{sf}A^{-1/3} + K_{\tau}I^2 + K_{coul}\frac{Z^2}{A^{4/3}} 
\quad ,
\eeqy
in which
\beqy\label{3.7}
K_{coul} = \frac{3e^2}{5r_0}\left(\frac{K^\prime}{K_v} - 8\right) \quad .
\eeqy
The parametrization (\ref{3.5}), which assumes a scaling model for the 
resonance, is based on the leptodermous approximation~\cite{dm69} and makes use
of the expansions (\ref{8a}) and (\ref{8b}) for infinite nuclear matter, only 
the lowest-order terms beyond $K_v$ being retained. Likewise, only the lowest
order associated with semi-infinite nuclear matter, $K_{sf}A^{-1/3}$, is 
retained. 

The value of $K_\tau$ that Ref.~\cite{lietal} extracts from the measurements 
is -550 $\pm$ 100 MeV, in clear-cut disagreement with all our forces 
(line 8 of Table~\ref{tab4}), and with the {\it ab initio} calculations of 
Ref.~\cite{vid09}. It also barely overlaps with the measurement of 
Ref.~\cite{chenetal}.  Moreover, it has been argued~\cite{piek09a,piek09b} 
that a value $K_{\tau} \approx$ -550 MeV is too strongly 
negative to be compatible with the behavior of low-density neutron matter, 
which is determined unambiguously by low-energy neutron-neutron scattering. 

However, as we have recently pointed out~\cite{pea10}, the analysis that 
Ref.~\cite{lietal} made of their data is invalid, essentially because as soon
as higher-order terms, notably the surface-symmetry term 
$K_{ss}I^2\,A^{-1/3}$, are admitted into the leptodermous expansion of 
$K(A,Z)$ then $K_{\tau}$ becomes indeterminate, and it is impossible to say 
that values in the vicinity of -350 MeV are not consistent with the data 
(see also Col\`o~\cite{col09}). Actually, the situation becomes even worse when
one realizes that it is not possible to identify the $K_v$ of Eq.~(\ref{3.5}) 
with the incompressibility of infinite nuclear matter $K_\infty$ if the 
structure of the breathing mode deviates from a strict scaling 
model~\cite{jj80,bla95}; similar considerations must also apply to $K_{\tau}$
itself, since it refers, like $K_v$, to a volume term. The only reliable way 
to see whether or not our forces are consistent with the breathing-mode data
would be to use them in self-consistent QRPA (or constrained HFB) calculations
of the breathing-mode energies~\cite{bla95}, which is beyond the scope of the
present article.

This approach has, in fact, been followed by Piekarewicz~\cite{piek07,piek09b}, 
whose RPA calculations are based on the relativistic mean-field method, using 
the FSU Gold parameters~\cite{piek05}. This
interaction reproduces well a certain number of key data points, and in
particular the measured breathing-mode energies of $^{208}$Pb, $^{144}$Sm, 
and $^{90}$Zr. On the other hand, his calculated breathing-mode energies
for the Sn isotopes lie significantly higher than the measured values of 
Ref.~\cite{lietal}. The value of $K_{\tau}$ for FSU Gold is -276.8 MeV, very 
close to our value for interaction BSk20, and differing insignificantly from
the values for the two other forces of this article. 

The two attempts that have been made to account for the anomalously soft 
properties of the Sn isotopes are summarized in Ref.~\cite{piek09b}, and found
to be incomplete: we are left with an open problem in nuclear structure.
As far as the present article is concerned the situation is rather troubling,
since as long as one does not understand the compressional properties of the 
Sn isotopes one cannot claim to fully understand those of the other nuclei 
from which values of $K_v$ are extracted. The
most that we can conclude is that at the present time the breathing-mode
measurements offer no basis for eliminating any of the forces discussed
here. On the other hand, if it turns out that the solution
to the tin problem lies in taking a value for $K_v$ that differs radically from 
what we have assumed here the present forces would all have to be abandoned 
and new fits made. 

\section{Landau parameters and stability conditions}
\label{llandau}

The values of the dimensionless Landau parameters in SNM at the equilibrium 
density $\rho_0$ are given in lines 12 -- 19 of 
Table~\ref{tab4}. All four forces predict almost identical values of 
the parameters $F_0$, $F_0^\prime$ and $F_1$, even though these parameters 
can in principle be modified by the new $t_4$ and $t_5$ terms (except in the 
case of force BSk18, where the conditions~(\ref{3a}) - (\ref{3c})
are imposed). The close agreement between the different forces can be traced 
back to the constraints imposed during the mass fit on the incompressibility 
$K_v$, the isoscalar effective mass $M_s^*$ and the symmetry energy $J$, since
these quantities are related uniquely to the Landau parameters through 
\bmlet
\beqy
\label{land.1a}
K_v=\frac{3\hbar^2 k_F^2}{M_s^*}(1+F_0)  \quad ,
\eeqy
\beqy
\label{land.1b}
\frac{M_s^*}{M}=1+\frac{F_1}{3}
\eeqy
and
\beqy
\label{land.1c}
J=\frac{\hbar^2 k_F^2}{6 M_s^*}(1+F_0^\prime)\quad .
\eeqy
\emlet

However the four forces yield quite different predictions for the remaining 
Landau parameters. The differences in the values of $F_1^\prime$ can easily
be understood in terms of the different values of the isovector effective 
mass, since the two quantities are related by
\beqy
\label{land.2}
\frac{M_s^*}{M_v^*}=1+\frac{F_1^\prime}{3}\, .
\eeqy

As for the parameters $G_0, G_0^\prime, G_1$, and $G_1^\prime$, it will be
seen that the three new forces give values that differ substancially from 
those of the older BSk18 force. In particular, $G_1$ and $G_1^\prime$ are now 
identically zero, while $G_0$ and $G_0^\prime$ are significantly higher.
This is a result of dropping the time-odd terms associated with the neglect of 
the $J^2$ and $J_q^2$ terms (see Sec~\ref{2A}). Our value of $G_0^\prime$, 
0.95 for all three new forces, is in good agreement 
with the value of around 1.2 found in BHF calculations using the Argonne A18 
potential with a three-body force~\cite{zuo03}. (On the other hand, all our 
values of $G_0$ are significantly lower than the value of around 0.8 found in 
these same realistic calculations, although some improvement over BSk18 can be 
seen in the three new models.) Our value of $G_0^\prime$ also falls within the 
empirical range of $1.0\pm0.1$ deduced in Ref.~\cite{bor84} from the analysis 
of Gamow-Teller (GT) resonances and magnetic-dipole modes in finite
nuclei. 

It is well-known that for the Fermi liquid theory to be internally 
consistent, the Landau parameters have to satisfy the two basic sum rules which 
follow from the Pauli exclusion principle~\cite{frim79}
\bmlet
\beqy
S_1=\sum_\ell \frac{F_\ell}{1+F_\ell/(2\ell+1)}+
\frac{F^\prime_\ell}{1+F^\prime_\ell/(2\ell+1)}  \nonumber \\
+\frac{G_\ell}{1+G_\ell/(2\ell+1)}+
\frac{G^\prime_\ell}{1+G^\prime_\ell/(2\ell+1)}  
=0   \quad ,
\eeqy
and
\beqy
S_2=\sum_\ell \frac{F_\ell}{1+F_\ell/(2\ell+1)}-
3\frac{F^\prime_\ell}{1+F^\prime_\ell/(2\ell+1)}   \nonumber \\
-3\frac{G_\ell}{1+G_\ell/(2\ell+1)}+
9\frac{G^\prime_\ell}{1+G^\prime_\ell/(2\ell+1)} 
=0     \quad .
\eeqy
\emlet
Skyrme forces generally violate these sum rules. 
However, we have found that our new forces significantly improve the second sum rule compared to 
our previous forces. In particular, while for BSk18 we have $S_2=3.05$, our new forces 
BSk19, BSk20 and BSk21 yield -0.13, -0.26 and -0.03, respectively. In the latter case, the 
discrepancy between the calculated value of $S_2$ and the exact result is reduced by 
two orders of magnitude. On the other hand, the values of the first sum $S_1$ for the different forces 
remain very close, lying between 0.7 and 0.9. All the forces quoted above include $t_4$ and $t_5$
terms. For comparison, our parametrization BSk17 based on a conventional Skyrme force yields
$S_1=-0.76$ and $S_2=7.42$. 

\begin{figure*}
\vskip -0.5cm
\centerline{\epsfig{figure=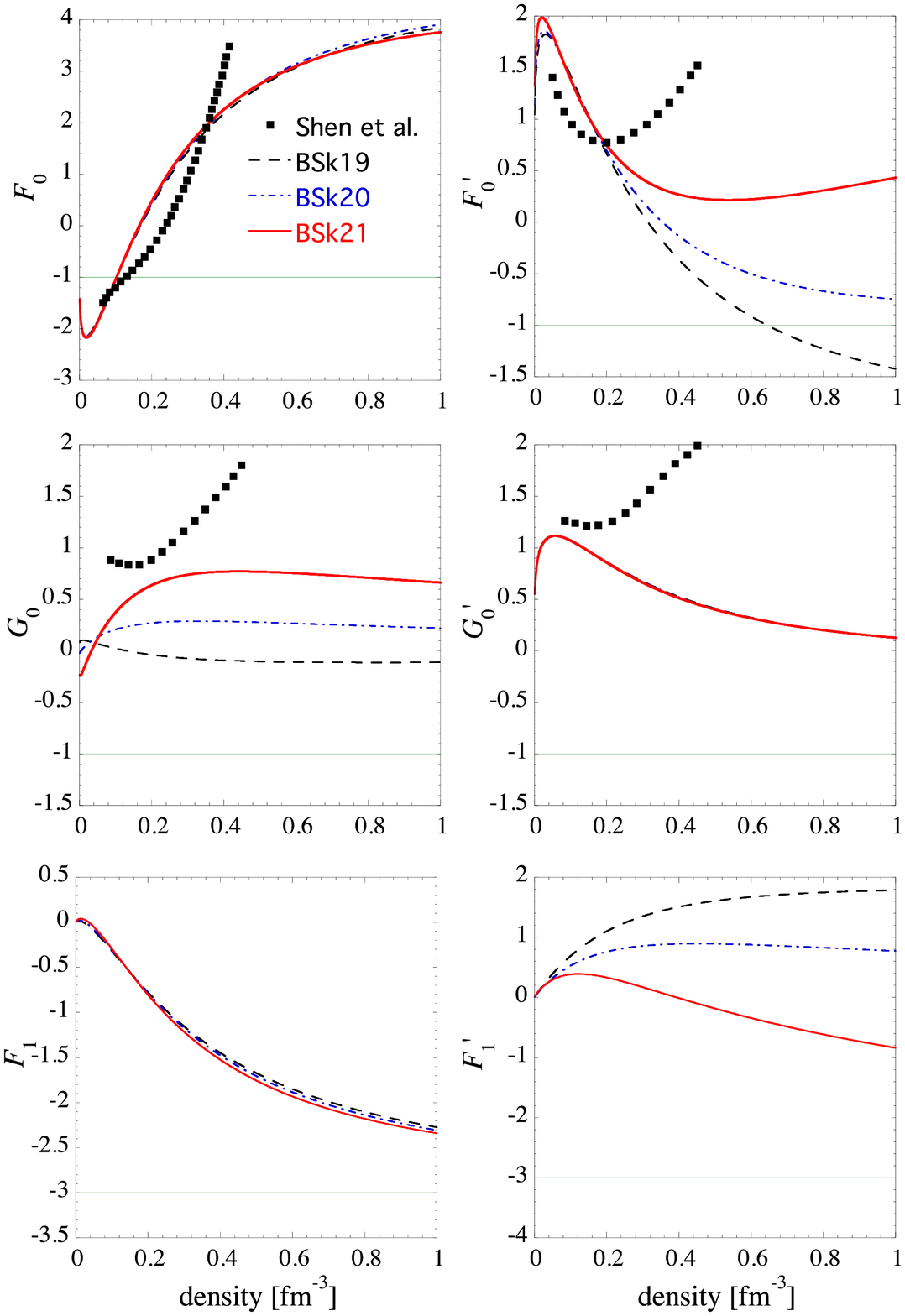,height=8.0cm}}
\caption{(Color online) Landau parameters in symmetric nuclear matter for 
forces BSk19 -- BSk21 (note that the curves for the three forces are
indistinguishable from each other in 
the case of $F_0, F_1$ and $G_0^\prime$). For comparison we show also the 
results of BHF calculations from Ref.~\cite{she03}. We also indicate the
stability limit $-(2\ell + 1)$.}
\label{figLanSNM}
\end{figure*}

Extending the calculation of the Landau parameters to higher densities we 
find, with one exception, that for all densities $\rho > \rho_0$ encountered in
the interior of supernova cores and neutron stars, the new forces satistify the 
following stability conditions for $\ell$ = 0 and 1, the only values 
applicable in the case of Skyrme forces:
\bmlet 
\beqy\label{staba}
 F_\ell > -(2\ell+1)\quad , \hskip0.5cm F_\ell^\prime > -(2\ell+1)
\eeqy
and
\beqy\label{stabb}
 G_\ell > -(2\ell+1)\quad ,\hskip0.5cm G_\ell^\prime > -(2\ell+1)\quad , 
\eeqy
\emlet
as shown in Figures~\ref{figLanSNM} (these conditions are still 
fulfilled when the time-odd terms associated with $J^2$ and $J_q^2$ terms are 
taken into account). 
The $\ell$ = 0 cases are of particular interest, as we now discuss.

(i) The condition on $G_0$ ensures that the spin asymmetry coefficient, defined 
by 
\beqy
a_\sigma \equiv \left(\frac{1}{2}
\frac{\partial^2 e}{\partial I_\sigma^2}\right)_{I_\sigma = 0} = 
\frac{\hbar^2 k_F^2}{6 M_s^*}(1+G_0)\quad ,
\eeqy
where $I_\sigma=(\rho_\uparrow-\rho_\downarrow)/\rho$, always remains positive,
i.e., SNM is stable with respect to a ferromagnetic transition. 

(ii) The condition on $G_0^\prime$ ensures that the spin-isospin 
asymmetry coefficient, defined by
\beqy
a_{\sigma\tau} \equiv \frac{1}{2}\left(\frac{\partial^2 e}{\partial I_{\sigma\tau}^2}\right)_{I_{\sigma\tau}=0}=\frac{\hbar^2 k_F^2}{6 M_s^*}(1+G_0^\prime)\quad ,
\eeqy
where $I_{\sigma\tau}=(\rho_{n\uparrow}-\rho_{n\downarrow}-
\rho_{p\uparrow}+\rho_{p\downarrow})/\rho$, always remains positive, i.e.,
there is no instability with respect to spin-isospin transitions.
Unlike our three new forces, BSk19, BSk20 and BSk21, our earlier model force 
BSk18 predicts that such an instability occurs at density of 
$\rho=0.62$ fm$^{-3}$. 

(iii) The condition on $F_0$ ensures that the incompressibility 
coefficient, defined by
\beqy
a_{\rho} \equiv \frac{1}{2}\frac{\partial^2 e}{\partial (\rho/\rho_0)^2} =
\frac{\hbar^2 k_F^2}{6 M_s^*}(1+F_0)\quad ,
\eeqy
is always positive, at least for $\rho > \rho_0$, i.e., SNM is stable against 
mechanical
breakup, of the sort that occurs for $\rho < \rho_0$. [Note that 
$a_{\rho}(\rho = \rho_0) = K_v/18\rho_0^2$, whence Eq.~(\ref{land.1a}) 
follows.]

(iv) The condition on $F_0^\prime$, where it holds, ensures that the isospin 
asymmetry coefficient, defined by
\beqy
a_{\tau} \equiv \frac{1}{2}\left(\frac{\partial^2 e}{\partial I_{\tau}^2}\right)_{I_\tau=0} = 
\frac{\hbar^2 k_F^2}{6 M_s^*}(1+F_0^\prime)\quad ,
\eeqy
where $I_{\tau} = (\rho_n - \rho_p)/\rho$, always remains positive, i.e.,
there is no isospin instability, in which the neutrons of SNM tend to become
protons, and {\it vice versa}. This holds at all densities $\rho > \rho_0$ for
forces BSk20 and BSk21, but breaks down for BSk19 above 
$\rho=0.64$ fm$^{-3}$, which is precisely the
density at which $e^{(1)}_{sym}$ goes negative for this force. (This is close
to, but not coincident with, the density at which $e^{(2)}_{sym}$ goes 
negative for BSk19, 0.73 fm$^{-3}$, as can be read off from Figure~\ref{fig3.2}.) 
There is a similar instability for BSk18.

In addition to instabilities in $F_0^\prime$ and $G_0^\prime$, BSk18 
predicts instability with respect to fluctuations in the spin-orbit current 
$\pmb{J}$, since $G_1$ falls below $-3$ at density $\rho=0.79$ fm$^{-3}$. 

\begin{figure*}
\vskip -0.5cm
\centerline{\epsfig{figure=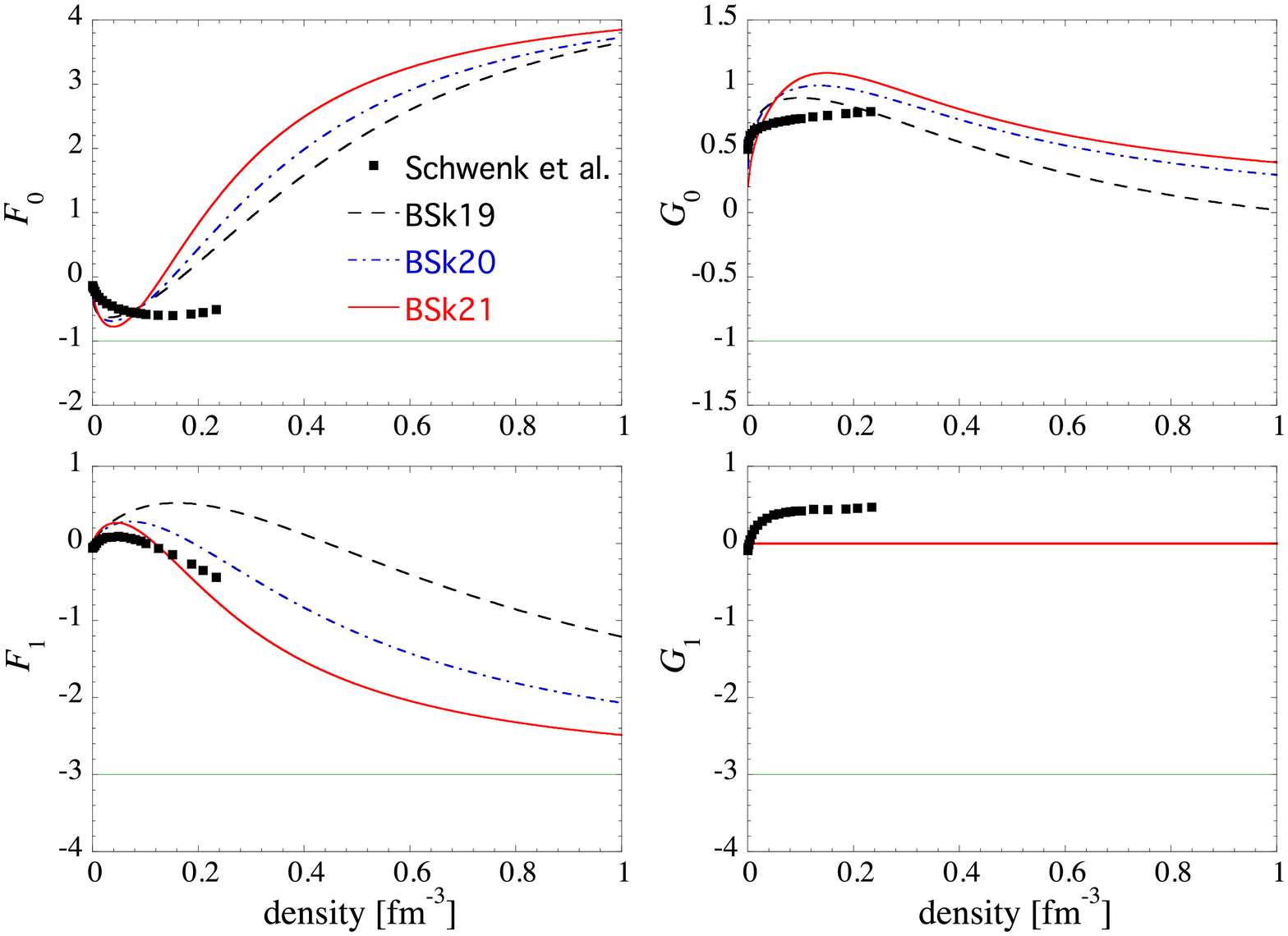,height=8.0cm}}
\caption{(Color online) Landau parameters in neutron matter for forces 
BSk19 -- BSk21 (note that the curves for the three forces are indistinguishable 
from each other in
the case of $G_1$). For comparison we show also the results of BHF 
calculations from Ref.~\cite{sfb03}. We also indicate the stability limit 
$-(2\ell + 1)$.}
\label{figLanNeuM}
\end{figure*}

We have also evaluated the Landau parameters in NeuM. The appropriate 
expressions for the generalized Skyrme force (\ref{2}) can be 
found in Appendix~\ref{landau}, while
the numerical values are shown in Figures~\ref{figLanNeuM}. As in the case 
of SNM, all the new forces BSk19 -- BSk21 are stable with respect to all the 
Landau parameters in NeuM. On the other hand, our earlier force 
BSk18 predicted an instability in $G_1^{\rm NeuM}$ at density 
$\rho=0.90$ fm$^{-3}$.

\section{Neutron-star matter}
\label{nsm}

We treat neutron-star matter (N*M), the matter constituting the homogeneous
core of neutron stars, as an electrically neutral $\beta$-equilibrated 
mixture of neutrons, protons, electrons and negative muons (muons were 
neglected in our article on mass model HFB-18 \cite{cgp09}).  Other particles, 
such as hyperons and quarks, may
appear toward the center of the star, but we neglect them here, since the 
many uncertainties associated with them would serve only to obscure the 
points that we want to make. 

\begin{figure}
\centerline{\epsfig{figure=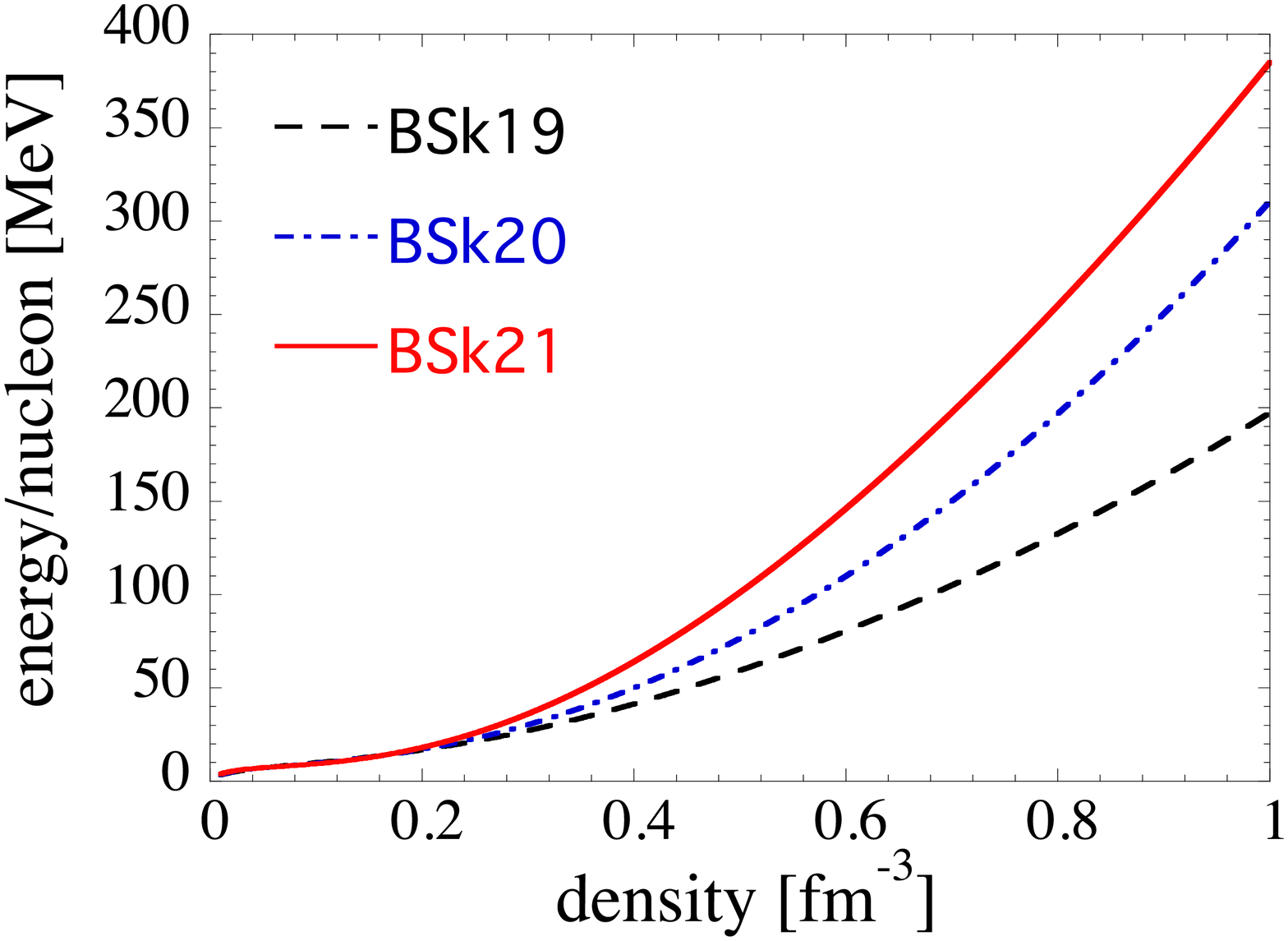,height=8.0cm}}
\caption{(Color online) EOSs of neutron-star matter for forces 
BSk19 -- BSk21.}
\label{fig4.1}
\end{figure}

\begin{figure}
\centerline{\epsfig{figure=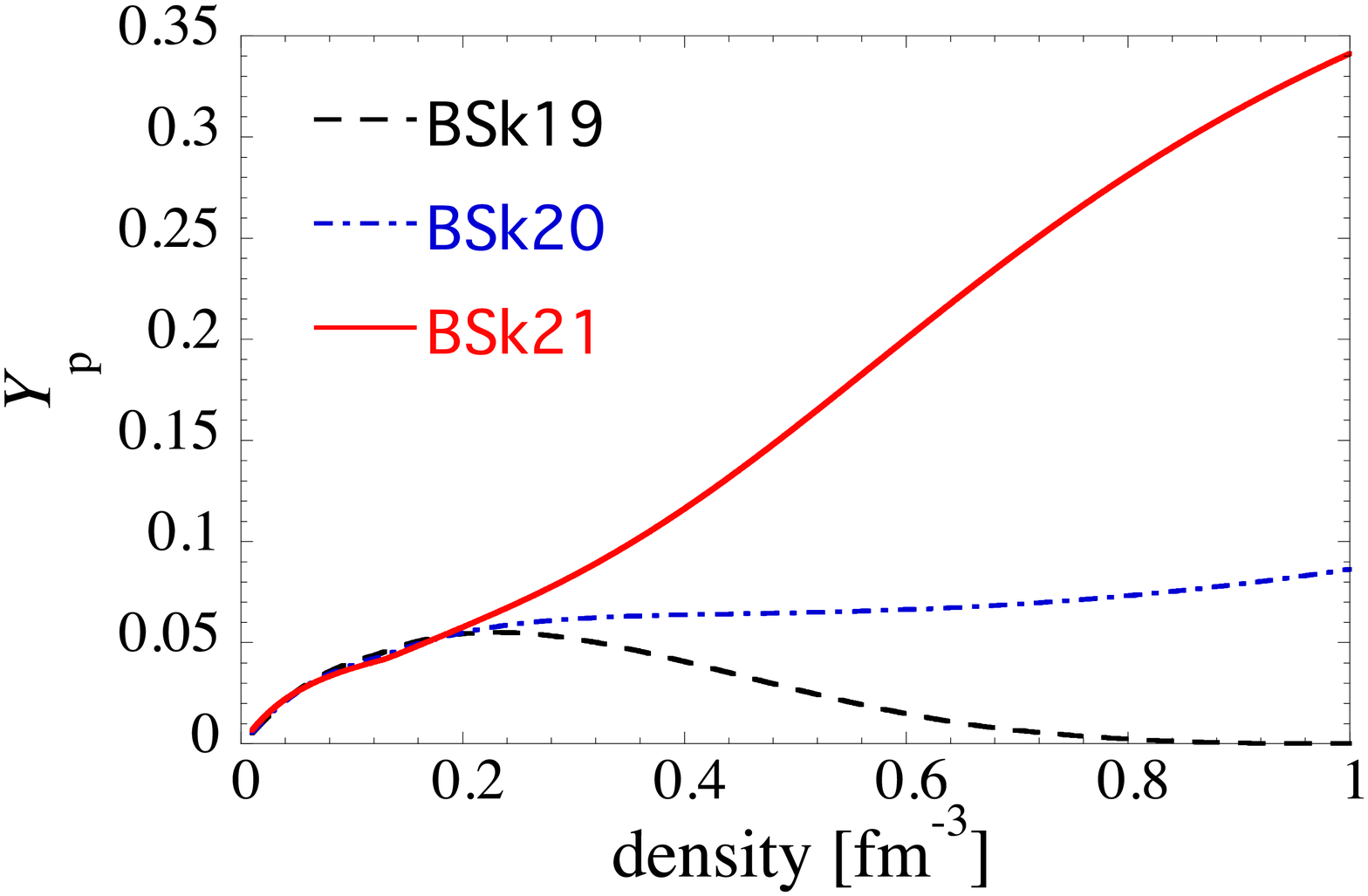,height=8.0cm}}
\caption{(Color online) Proton fraction $Y_p$ in neutron-star matter for 
forces BSk19 -- BSk21.}
\label{fig4.2}
\end{figure}

\begin{figure}
\centerline{\epsfig{figure=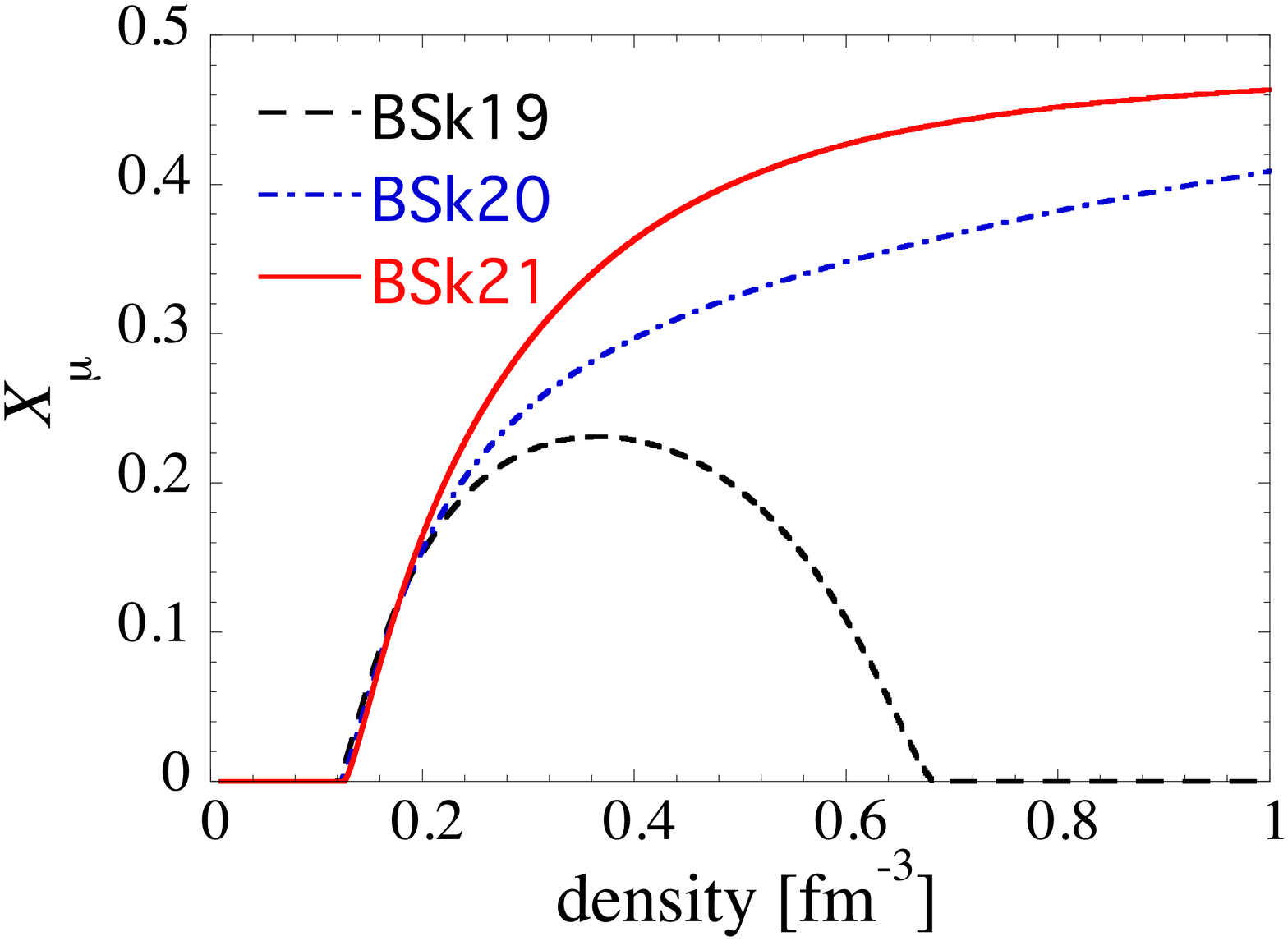,height=8.0cm}}
\caption{(Color online) Number of muons per proton in neutron-star matter for 
forces BSk19 -- BSk21}
\label{fig4.3}
\end{figure}

{\it Equation of state and composition.} Figure~\ref{fig4.1} shows the 
zero-temperature EOS of N*M for each of our three new forces (the EOS for the
older force, BSk18, is quite similar to that of BSk19). In each case we have, 
as mentioned earlier, checked that N*M remains unpolarized over 
the whole density range. 
In Figure~\ref{fig4.2} we see as a function of density the fraction $Y_p$ of 
nucleons that are protons, while Figure~\ref{fig4.3} shows the fraction 
$X_{\mu}$ of leptons that are muons (the number of leptons per nucleon is 
just $Y_p$).

Comparing Figure~\ref{fig1} with Figure~\ref{fig4.1}, we see that although the EOS 
of N*M might be considerably softer than the corresponding EOS of NeuM, there 
is still a correlation between the two systems in the sense that the stiffer 
the EOS of NeuM, the stiffer the EOS of N*M. Likewise, comparison of 
Figure~\ref{fig3.2} with Figure~\ref{fig4.2} shows that the symmetry energy 
$e^{(2)}_{sym}$ is strongly correlated with the protonic fraction $Y_p$, as
is well known (see, for example, Sec 5.11.2 of Ref.~\cite{hae07}). It is 
particularly to be noted that even though the EOS of BSk19 is ``supersoft",
in the sense that the symmetry energy turns negative at densities encountered
in neutron-star cores, it will still support a stable neutron star (contrary to
the suggestion of Ref.~\cite{wen09}), since, as
seen in Figure~\ref{fig4.1}, the energy per nucleon increases monotonically with 
density, i.e., the pressure is always positive. Indeed, precisely because the
EOS is ``supersoft", N*M becomes, for BSk19, pure NeuM at higher densities, and
the symmetry energy plays no direct role.
 
The value of $Y_p$ found at any given density in a neutron star is of 
considerable interest, since if, but only if, it exceeds $\approx$ 11-15\% 
a direct Urca process of neutrino cooling \cite{lat91} will
be possible. Inspection 
of Figure~\ref{fig4.2} tells us that with BSk19 a direct Urca process will be 
quite impossible, with BSk20 it will be unlikely, while it will certainly occur
for BSk21. We recall that the one available experimental result on the 
high-density behavior of the symmetry energy tends to favor the BSk19 force, 
thereby casting doubt on the possibility of a direct Urca process. On the
other hand, the low luminosity from the pulsar in CTA 1 and from several
young supernova remnants likely to contain a still unobserved neutron
star~\cite{kapl04,kapl06} might be an indication that a direct Urca
process was actually occurring~\cite{sy08, pa09}. 
Evidently, an enormous amount of work, both theoretical and experimental, 
remains to be done on the question of symmetry energy at the high densities 
found toward the center of neutron stars. In the meantime, a clear
demonstration that a direct Urca process was or was not contributing to 
neutron-star cooling would serve as a most valuable signpost to nuclear 
physicists. 

{\it Causality.} A necessary condition for the validity of our forces is
that the speed of sound $v_s$ in N*M must not exceed the speed of light $c$
at the densities encountered in neutron stars.
We have therefore calculated the density $\rho_c$ above which this condition is
violated. The velocity of sound is given by (see Sec 5.13.3 of 
Ref.~\cite{hae07})
\beqy
\frac{v_s}{c} = 
\sqrt{\left(\frac{{\rm d}\,P}{{\rm d}\,\mathcal{E}}\right)_{\rm fr}}=
\sqrt{\frac{\gamma_{\rm fr} P}{\mathcal{E}+P}} \quad  ,
\eeqy
in which $P$ is the pressure, $\mathcal{E}$ is the total 
energy density (including the rest-mass energy) and $\gamma_{\rm fr}$ is the 
adiabatic index, defined by
\beqy
\gamma_{\rm fr} = 
\frac{\rho}{P}\left(\frac{{\rm d}P}{{\rm d}\rho}\right)_{\rm fr} \quad ;
\eeqy
the subscript ``fr" is to indicate that the derivatives have to 
evaluated with the composition frozen.
The value of $\rho_c$ for the different forces are given in Table~\ref{tab4}. 
Even though the contribution of electrons and muons to the EOS is rather small, 
it was included in $v_s$, treating them as relativistic Fermi gases. 
As can be seen from Figure~\ref{fig4.1}, the stiffer the EOS, the 
lower $\rho_c$ is. For comparison we have also shown in Table~\ref{tab4} the 
critical density $\rho_c$ in pure NeuM. For force BSk19 the critical
densities in NeuM and in N*M are the same because at high densities N*M 
consists of neutrons only. On the other hand, for BSk21 N*M contains a sizable 
amount of proton-lepton pairs, their effect on the EOS and $v_s$ being to
significantly increase $\rho_c$. For all our forces the relatively low values 
of $\rho_c$(N*M) that we find may lead to a violation of causality in heavier 
neutron stars, but in this respect we are limited by the corresponding 
violation that occurs in the realistic EOSs of NeuM to
which we have fitted our forces.

\section{Conclusions}
\label{concl}

This article describes the latest effort in our long-standing quest
for effective forces (Skyrme plus contact pairing) that will lead to 
high-precision mass models, while at the same time respecting the 
physical constraints appropriate to the neutron-rich environments found in 
neutron stars and supernova cores. Specifically, we present here a family 
of three generalized Skyrme forces, each of which, when taken with the 
appropriate pairing 
force, leads to a mass model that fits essentially all the mass 
data~\cite{audi03} with an rms deviation of about 0.58 MeV. However, although
these models give very similar extrapolations out to the neutron drip line, 
the corresponding Skyrme forces are distinguished by the very different way in 
which the symmetry energy of each varies at the high densities found in the 
core of neutron stars, ranging from the supersoft (i.e., negative
symmetry energy) to the very stiff. This degree of flexibility, maintaining
a high quality mass fit with very different neutron-matter constraints, has 
been made possible only through the introduction of the $t_4$
and $t_5$ terms, i.e., density-dependent generalizations of the usual $t_1$ and
$t_2$ terms, respectively. Despite these differences, each of these 
Skyrme forces finds theoretical support in the sense of having been fitted to 
one realistic many-body calculation or another of neutron matter. Moreover, 
the neutron matter corresponding to each of these forces is stable against
unphysical transitions, such as a ferromagnetic flip. Finally, in symmetric
nuclear matter the distribution of potential energy among the different
spin-isospin channels is in qualitative accord with realistic many-body 
calculations. 

We claim now that with the forces presented here, along with their respective 
mass models, it has become possible for the first time to adopt a unified
treatment at the level of effective forces of all the nuclear properties of the
highly neutron-rich systems of astrophysical interest. Thus by constraining the
forces to fit not only all the available mass data but also the EOS of
neutron matter our HFB mass models may be expected to provide the most
reliable estimates of the masses of the experimentally inaccessible highly
neutron-rich nuclei that appear in the outer crust of neutron stars and that 
are involved in the $r$-process (note that the predictions of our three forces 
for neutron matter begin to diverge from each other only at densities of about 
2$\rho_0$).  Moreover, since our forces, having been fitted not only to neutron
matter but also to nuclear masses, take account
of both inhomogeneities and the presence of protons they can be used with
confidence to calculate the EOS of the inner crust of neutron stars. 
Indeed, since our forces do not lead to any unphysical instability in neutron 
matter they can likewise be used for the study of the core of neutron stars,
whence a unified treatment of the whole star becomes possible. 

Of course, the three different Skyrme forces will make quite different 
predictions for the properties of the core (and thus for global properties
such as the mass - radius relationship), but this simply reflects our present
ignorance of the properties of neutron matter at high density. But even if
many-body theory could be developed to the point where the EOS of neutron 
matter could be tied down unambiguously, there would still be uncertainties
associated with the possible presence of hyperons and other exotica. Thus
we believe that the ultimate discrimination among our different forces will
come either from experiment or observation of neutron stars. (As for
experiment, we stress that measurement must relate to high densities: we cannot
draw conclusions about the high-density behavior 
from low-density quantities such as the symmetry-compressibility coefficients 
$K_{sym}$ or $K_{\tau}$, even if measured with sufficient precision.)
But whichever of our three forces is ultimately favored, we have shown here 
that the Skyrme form of effective interaction is flexible enough to permit a
description of the 
nuclear physics of the entire neutron star with a unique set of parameters.

We also want to be able to apply the same Skyrme force to the calculation 
of nuclear quantities relevant to the $r$-process. In addition to masses, such 
quantities include beta strength functions, fission barriers and level 
densities. In connection with the first of these quantities an outstanding 
difficulty with all our previously published HFB models has been that of
obtaining a reasonable value for the Landau parameter $G_0^\prime$; for
the first time we can claim that with our forces BSk19 -- BSk21 we have
resolved this problem. Previous articles have already dealt successfully
with fission barriers~\cite{sg07} and level densities~\cite{sg06,hg06}. 

\begin{acknowledgments}
We thank M. Baldo, U. Garg and U. Lombardo for valuable exchanges. 
This work was financially supported by the FNRS (Belgium), the Communaut\'e 
fran\c{c}aise de Belgique (Actions de Recherche Concert\'ees), CompStar (a 
Research Networking Programme of the European Science Foundation) and
the NSERC (Canada). J. M. P. is grateful to Bureau
des relations internationales of the Universit\'e Libre de Bruxelles
for financial support for the month of November 2009.
\end{acknowledgments}

\appendix

\section{Microscopic calculation of collective quadrupole energy}
\label{gognycoll}

\begin{figure}
\centerline{\epsfig{figure=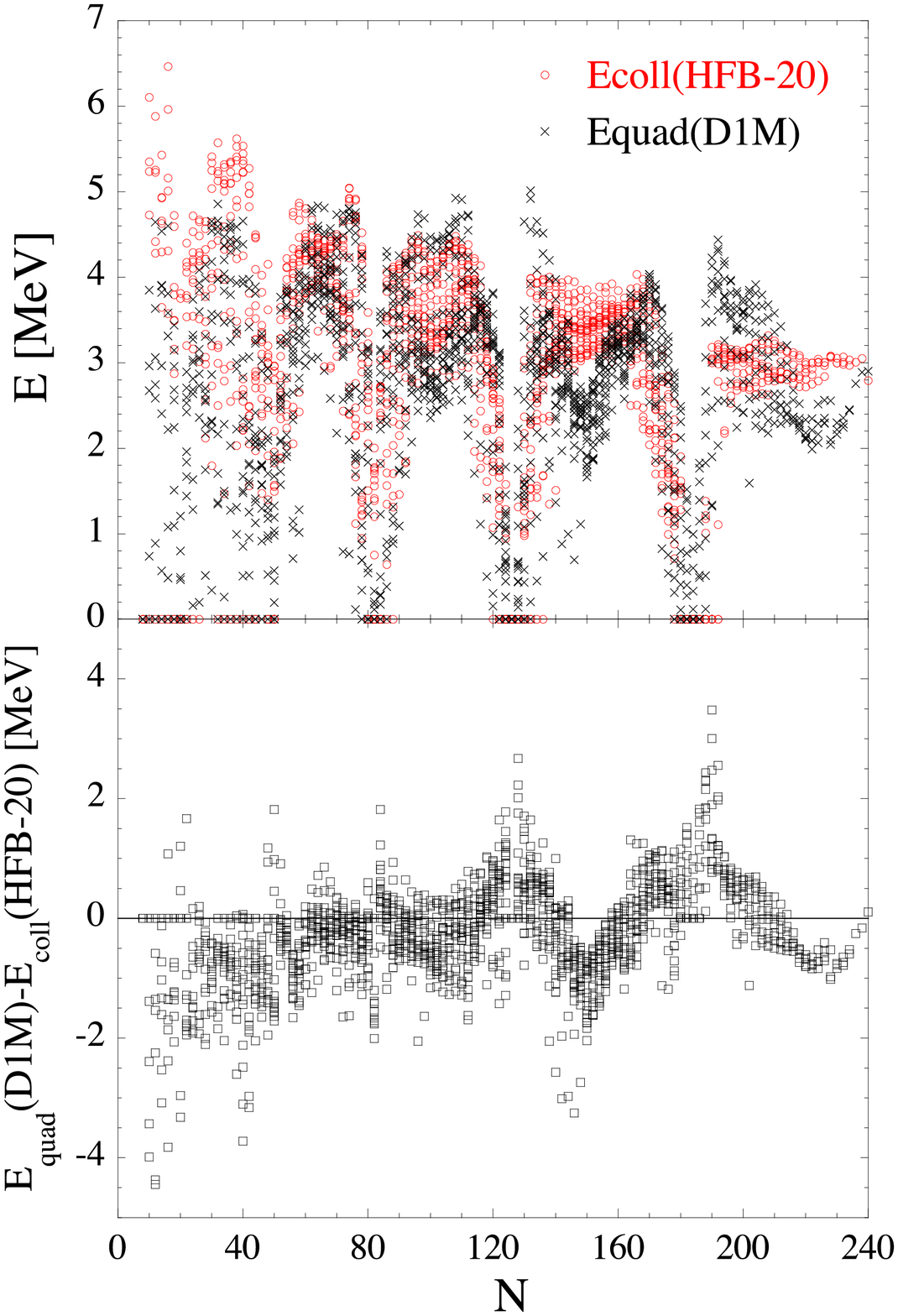,height=8.0cm}}
\caption{(Color online) (Upper panel) Comparison between the D1M quadrupole 
correction energy (crosses) \cite{gg09} and the HFB-20 collective energy 
(circles) as a function of $N$ for the 1770 even-even nuclei with $Z,N > 8$ and
$Z\le 110$. (Lower panel) Differences between the D1M quadrupole correction 
energy and the HFB-19 collective energy.}
\label{fig_equad}
\end{figure}

Systematic calculations of the collective quadrupole energy have recently 
been performed on the basis of the D1M Gogny interaction by solving the 
collective Schr\"odinger equation with the five-dimensional collective Hamiltonian
\cite{gg09}. The five collective quadrupole coordinates correspond to the 
rotation, the quadrupole vibration and the coupling between these collective 
modes; it includes axial as well as triaxial quadrupole deformations. The 
corresponding quadrupole correction energy obtained with the D1M force is 
compared in Figure~\ref{fig_equad} with the collective correction 
energy~(\ref{6}) for the 
1770 even-even nuclei with $Z,N > 8$ and $Z\le 110$. Globally, our simple 
approximation is seen to be in relatively good agreement with the D1M 
quadrupole energies, so we can be confident that most of the quadrupole 
effects are indeed explicitly treated and not taken at the level of the mean 
field calculation (hence not absorbed into 
the effective force). Some systematic 
deviations between both approaches can be observed in the lower panel of 
Figure~\ref{fig_equad}. Interestingly these differences are correlated to those 
found between experimental and D1M masses (see in particular Figure~1 of 
Ref.~\cite{gg09}) for $N > 60$ nuclei, so mid-shell correlations, in 
particular around $N\simeq 100$ and 150,  may in fact be underestimated by the 
collective Hamiltonian model.

\section{Landau parameters in neutron matter}
\label{landau}

The expressions of the Landau parameters for symmetric nuclear matter
associated with generalized Skyrme forces~(\ref{2}) were given in
Ref.~\cite{cgp09}. Here we provide the expressions of the Landau parameters
in pure neutron matter, using the same notations.

\bmlet
\beqy
F_0^{\rm NeuM}=N \Biggl[2 C_0^\rho+2C_1^\rho+2 k_{\rm F}^2(C_0^\tau+C_1^\tau) +4\rho\left( \frac{{\rm d} C_0^\rho}{{\rm d}\rho}+ \frac{{\rm d} C_1^\rho}{{\rm d}\rho}\right)\nonumber \\ +\rho^2\left(\frac{{\rm d}^2 C_0^\rho}{{\rm d}\rho^2}+\frac{{\rm d}^2 C_1^\rho}{{\rm d}\rho^2}\right) +\rho\tau\left(\frac{{\rm d}^2 C_0^\tau}{{\rm d}\rho^2}+\frac{{\rm d}^2 C_1^\tau}{{\rm d}\rho^2}\right)+\left(\frac{{\rm d} C_0^\tau}{{\rm d}\rho}+\frac{{\rm d} C_1^\tau}{{\rm d}\rho}\right)(2\tau+2\rho k_{\rm F}^2)\Biggr]   \quad ,
\eeqy
\beqy
F_1^{\rm NeuM}=-2 N (C_0^\tau+C_1^\tau) k_{\rm F}^2    \quad ,
\eeqy
\beqy
G_0^{\rm NeuM}=2N\Biggl[C_0^s+C_1^s+ k_{\rm F}^2(C_0^T+C_1^T)\Biggr]
\eeqy
and
\beqy
G_1^{\rm NeuM}=-2 N k_{\rm F}^2 (C_0^T+C_1^T)   \quad ,
\eeqy
\emlet
where $N$ is the density of s.p. states at the Fermi level
\beqy
N=\frac{M^*_n k_{\rm F}}{\hbar^2\pi^2}\, ,
\eeqy
$k_{\rm F}=(3\pi^2\rho)^{1/3}$
and $M_n^*$ is the neutron effective mass given by
\beqy
\frac{M_n}{M_n^*}=1+\frac{2M_n}{\hbar^2}(C_0^\tau+C_1^\tau)\rho\, .
\eeqy

\begin{table}
\centering
\caption{Forces BSk19 -- BSk21: lines 1-16 show the Skyrme parameters, 
lines 17-21 the pairing parameters, and the last 4 lines the Wigner 
parameters (see text for further details). Note that it is more convenient to 
show the $x_2$ parameter in the form $t_2x_2$, the only combination in which 
$x_2$ enters into the formalism. For convenience of comparison we also show
the force BSk18 \cite{cgp09}.}
\label{tab1}
\vspace{.5cm}
\begin{tabular}{|c|ccc|c|}
\hline
  &BSk19&BSk20&BSk21&BSk18\\
\hline
  $t_0$ {\scriptsize [MeV fm$^3$]}&-4115.21 &-4056.04&-3961.39  & -1837.96  \\
  $t_1$ {\scriptsize [MeV fm$^5$]}&403.072 & 438.219&396.131  & 428.880   \\
  $t_2$ {\scriptsize [MeV fm$^5$]}&0 &0&0 & -3.23704\\
  $t_3$ {\scriptsize [MeV fm$^{3+3\alpha}$]}&23670.4&23256.6&22588.2  & 11528.9  \\
  $t_4$ {\scriptsize [MeV fm$^{5+3\beta}$]}&-60.0 &-100.000 &-100.000& -400.000
\\
  $t_5$ {\scriptsize [MeV fm$^{5+3\gamma}$]}&-90.0&-120.000&-150.000  & -400.000
\\
  $x_0$          &0.398848     & 0.569613 &0.885231          &  0.421290  \\
  $x_1$   &-0.137960  &-0.392047 &0.0648452  & -0.907175\\
  $t_2x_2$ {\scriptsize [MeV fm$^5$]}&-1055.55 &-1147.64 &-1390.38 & -186.837 \\
  $x_3$  &0.375201           &0.614276&1.03928 &  0.683926 \\
  $x_4$ &-6.0                  &-3.00000 &2.00000          & -2.00000  \\
  $x_5$  &-13.0                 &-11.0000 &-11.0000          &  -2.00000 \\
  $W_0$ {\scriptsize [MeV fm$^5$]}&110.802&110.228&109.622   &  138.904    \\
  $\alpha$   &1/12                &1/12 &1/12       &  0.3  \\
  $\beta$     &1/3               &1/6&1/2&1.0  \\
  $\gamma$    &1/12               &1/12&1/12        &1.0 \\
  $f_{n}^+$&1.00 & 1.00&1.00&  1.00 \\
  $f_{n}^-$&1.05 &1.06 &1.05&  1.06  \\
  $f_{p}^+$&1.10 &1.09 &1.07&  1.04  \\
  $f_{p}^-$&1.17 &1.16 &1.13&  1.09   \\
  $\varepsilon_{\Lambda}$ {\scriptsize [MeV]} &16.0&16.0&16.0&  16.0    \\
  $V_W$ {\scriptsize [MeV]}&-2.00&  -2.10 & -1.80   & -2.10   \\
  $\lambda$    &250  &280           &     280     & 340    \\
  $V_W^{\prime}$ {\scriptsize [MeV]}&1.16&0.96&0.96  & 0.74   \\
  $A_0$         & 24   & 24     & 24   &28   \\
 \hline
\end{tabular}
\end{table}

\begin{table}
\centering
\caption{Parameters of Eq.(\ref{6}) for collective correction to models
HFB-19, HFB-20 and HFB-21.}
\label{tab2}
\vspace{.5cm}
\begin{tabular}{|c|c|}
\hline
$b$  (MeV)& 0.80 \\
$c$ &10  \\
$d$ (MeV) & 3.4 \\
$l$ & 17 \\
$\beta_2^0$ &0.1\\
 \hline
\end{tabular}
\end{table}

\begin{table}
\centering
\caption{Rms ($\sigma$) and mean ($\bar{\epsilon}$) deviations between data
and predictions for models HFB-19, HFB-20 and HFB-21; for convenience of 
comparison we also show model HFB-18 \cite{cgp09}. 
The first pair of lines refers to all the 2149 measured masses $M$ that were 
fitted \cite{audi03}, the second pair to the masses $M_{nr}$ of the subset of
185 neutron-rich nuclei with $S_n \le $ 5.0 MeV, the third pair to the 
neutron separation energies $S_n$ (1988 measured values), the fourth pair to 
beta-decay energies $Q_\beta$ (1868 measured values) and the fifth pair to 
charge radii (782 measured values \cite{ang04}). The last line shows
the calculated neutron-skin thickness of $^{208}$Pb for these models.}
\label{tab3}
\vspace{.5cm}
\begin{tabular}{|c|ccc|c|}
\hline
&HFB-19&HFB-20&HFB-21&HFB-18  \\
\hline
$\sigma(M)$ {\scriptsize [MeV]}&0.583  &0.583 &0.577 &0.585  \\
$\bar{\epsilon}(M)$ {\scriptsize [MeV]}&-0.038 &0.021 &-0.054&0.007 \\
$\sigma(M_{nr})$ {\scriptsize [MeV]}&0.803 &0.790 &0.762&0.758  \\
$\bar{\epsilon}(M_{nr})$ {\scriptsize [MeV]}&0.243 &0.217 &-0.086& 0.172\\
$\sigma(S_n)$ {\scriptsize [MeV]}&0.502 &0.525&0.532 &0.487 \\
$\bar{\epsilon}(S_n)$ {\scriptsize [MeV]}&-0.015 &-0.012 &-0.009 &  -0.012\\
$\sigma(Q_\beta)$ {\scriptsize [MeV]}&0.612 &0.620&0.620&0.561   \\
$\bar{\epsilon}(Q_\beta)$ {\scriptsize [MeV]}&0.027 &0.024&0.000& 0.025\\
$\sigma(R_c)$ {\scriptsize [fm]}&0.0283&0.0274&0.0270&0.0274\\
$\bar{\epsilon}(R_c)$ {\scriptsize [fm]}&-0.0032&0.0009&-0.0014&0.0016\\
$\theta$($^{208}$Pb) {\scriptsize [fm]}&0.140&0.140&0.137&0.150\\
\hline
\end{tabular}
\end{table}

\begin{table}
\centering
\caption{Parameters of infinite nuclear matter for forces BSk19, BSk20 and 
BSk21; for convenience of comparison we also show force BSk18 \cite{cgp09}.} 
\label{tab4}
\vspace{.5cm}
\begin{tabular}{|c|ccc|c|}
\hline
&BSk19&BSk20&BSk21&BSk18\\
\hline
$a_v$ {\scriptsize [MeV]}&-16.078&-16.080&-16.053&-16.063 \\ 
$\rho_0$ {\scriptsize [fm$^{-3}$]}&0.1596&0.1596&0.1582&0.1586 \\
$J$ {\scriptsize [MeV]}&30.0&30.0  &30.0 &30.0 \\
$K_v$ \scriptsize [MeV]&237.3&241.4&245.8&241.8 \\
$K^{\prime}$ \scriptsize [MeV]&297.8&282.2&274.1&363.8\\
$L$ \scriptsize [MeV]&31.9&37.4  &46.6  &36.2    \\
$K_{sym}$\scriptsize [MeV]&-191.4&-136.5&-37.2&-180.9\\ 
$K_{\tau}$\scriptsize [MeV]&-342.8&-317.1&-264.6 &-343.7 \\ 
$K_{coul}$\scriptsize [MeV]&-5.093&-5.158&-5.186&-4.897 \\
$M^*_s/M$&0.80&0.80 &0.80&0.80 \\
$M^*_v/M$&0.61&0.65& 0.71 &0.79 \\ 
$F_0$&-0.14& -0.13 & -0.10 &-0.12\\
$F_0^{\prime}$&0.96& 0.96 & 0.97 &0.97\\
$F_1$&-0.60& -0.60 & -0.60 &-0.60\\
$F_1^{\prime}$&0.96& 0.69 & 0.38 &0.032\\
$G_0$ &-0.016& 0.25 & 0.56 &-0.33\\
$G_0^{\prime}$&0.95& 0.95 & 0.95 &0.46\\
$G_1$&0& 0 & 0 &1.23\\
$G_1^{\prime}$&0& 0 & 0 &0.50\\
$\rho_c$(N*M) {\scriptsize [fm$^{-3}$]}&1.45&0.98& 0.99 & 1.77 \\
$\rho_c$(NeuM) {\scriptsize [fm$^{-3}$]}&1.45&0.95& 0.69 & 1.77 \\
\hline
\end{tabular}
\end{table}

\begin{table}
 \centering
 \caption{Single-particle neutron levels in $^{208}$Pb (MeV). 
Experimental values are taken from Ref.\ \cite{vb72}. The asterisk denotes
the Fermi level. The quantity $\Delta_n$ is the interval between the centroids 
of the $2f$ and $3d$ doublets.} 
\label{tab5}
 \vspace{.5cm}
 \tabcolsep=.5cm
\vspace{0.5cm}
\begin{center}
\begin{tabular}{|c|cc|c|}
\hline
Level       &  HFB-18 &  HFB-20 & Expt.\\
\hline
$1s_{1/2}$  & -51.0 & -49.1  &- \\
\ldots & \ldots & \ldots &  \ldots \\
$1h_{9/2}$  & -12.6 & -11.7  &-10.9\\
$2f_{7/2}$  & -11.6 & -11.3  &-9.7\\
$1i_{13/2}$ &  -9.4 &  -9.7  &-9.0\\
$3p_{3/2}$  &  -8.8 &  -8.7  &-8.3\\
$2f_{5/2}$  &  -8.8 &  -8.6  &-8.0\\
$3p_{1/2}$* &  -7.7 &  -7.7  &-7.4\\
$2g_{9/2}$  &  -3.5 &  -3.7  &-3.9\\
$1i_{11/2}$ &  -2.7 &  -2.3  &-3.2\\
$1j_{15/2}$ &  -1.2 &  -1.8  &-2.5\\
$3d_{5/2}$  &  -1.2 &  -1.5  &-2.4\\
$4s_{1/2}$  &  -0.7 &  -0.9  &-1.9\\
$2g_{7/2}$  &  -0.4 &  -0.7  &-1.5\\
$3d_{3/2}$  &  -0.2 &  -0.5  &-1.4\\ 
\hline
$\Delta_n$  &9.6  &9.0   & 7.0\\
\hline
\end{tabular}
\end{center}
\end{table}

\begin{table}
\centering
 \centering
 \caption{Single-particle proton levels in $^{208}$Pb (MeV).
Experimental values are taken from Ref.\ \cite{vb72}. The asterisk denotes
the Fermi level. The quantity $\Delta_p$ is the interval between the centroids
of the $1g$ and $2f$ doublets.}
\label{tab6}
 \vspace{.5cm}
 \tabcolsep=.5cm
\vspace{0.5cm}
\begin{center}
\begin{tabular}{|c|cc|c|}
\hline
Level  &  HFB-18 &  HFB-20 & Expt.\\
\hline
$1s_{1/2}$  & -39.8 & -41.0  & -\\
\ldots & \ldots & \ldots  & \ldots\\
$1g_{9/2}$  & -16.3 & -16.6  &-15.4\\
$1g_{7/2}$  & -13.0 & -12.7  & -11.4\\
$2d_{5/2}$  &  -10.2 &  -10.3  &-9.7\\
$1h_{11/2}$ &  -8.8 &  -9.0  &-9.4\\
$2d_{3/2}$  &  -8.4 &  -8.6 &-8.4\\
$3s_{1/2}$* &  -7.7 &  -7.8  &-8.0\\
$1h_{9/2}$  &  -4.0&  -3.5  &-3.8\\
$2f_{7/2}$  &  -2.2 &  -2.2  &-2.9\\
$1i_{13/2}$ &  -1.0 &  -1.1  &-2.2\\
$3p_{3/2}$  &  1.0 &  0.9  &-1.0\\
$2f_{5/2}$  &  0.2 &  0.1  &-0.5\\
\hline
$\Delta_p$ &13.6  &13.7&11.7\\
\hline
\end{tabular}
\end{center}
\end{table}


\begin{thebibliography}{99}
\bibitem{gcp09}S. Goriely, N. Chamel, and J.~M. Pearson, Phys. Rev. Lett. 
{\bf 102}, 152503 (2009).
\bibitem{audi03} G. Audi, A.H. Wapstra, and C. Thibault,
Nucl. Phys. \textbf{A729}, 337  (2003).
\bibitem{onsi08}M. Onsi, A.~K. Dutta, H. Chatri, S. Goriely, N. Chamel, and
J.~M. Pearson, Phys. Rev. C {\bf 77} 065805 (2008).
\bibitem{cgp09}N. Chamel, S. Goriely, and J. M. Pearson, 
Phys. Rev. C {\bf 80}, 065804 (2009).
\bibitem{fp81}B. Friedman and V.~R. Pandharipande, Nucl. Phys. {\bf A361},
502 (1981).
\bibitem{apr98}A. Akmal, V.~R. Pandharipande, and D. G. Ravenhall,
Phys. Rev. C {\bf 58}, 1804 (1998).
\bibitem{xiao09} Z. Xiao, B.-A. Li, L.-W. Chen, G.-C. Yong, and M. Zhang,
Phys. Rev. Lett. {\bf 102}, 062502 (2009).
\bibitem{ls08}Z. H. Li and H.-J. Schulze,
Phys. Rev. C  \textbf{78}, 028801 (2008). 
\bibitem{gan09}S. Gandolfi, A. Yu. Illarionov, K. E. Schmidt, F. Pederiva,
and S. Fantoni, Phys. Rev. C  \textbf{79}, 054005 (2009).
\bibitem{cao06}L~.G. Cao, U.~Lombardo, and P.~Schuck,
Phys. Rev. C  \textbf{74}, 064301  (2006).
\bibitem{cgp08} N. Chamel, S. Goriely, and J. M. Pearson,
Nucl. Phys. {\bf A812}, 72 (2008).
\bibitem{sg02}S. Goriely, M. Samyn, P.-H. Heenen, J. M. Pearson, and
F. Tondeur, Phys. Rev. C  \textbf{66}, 024326  (2002).
\bibitem{sg05}S. Goriely, M. Samyn, J.~M. Pearson, and M. Onsi,
Nucl. Phys. {\bf A750}, 425 (2005).
\bibitem{bend04}M. Bender, P.-H. Heenen, and P. Bonche,
Phys. Rev. C {\bf 70}, 054304 (2004).
\bibitem{sg07}S. Goriely, M. Samyn, and J.M. Pearson,
Phys. Rev C {\bf 75}, 064312 (2007).
\bibitem{gp08}S. Goriely and J. M. Pearson,
Phys. Rev. C  \textbf{77}, 031301(R) (2008).
\bibitem{doba95}J. Dobaczewski and J. Dudek, Phys. Rev. C {\bf 52}, 1827
(1995).
\bibitem{bend02}M. Bender, J. Dobaczewski, J. Engel, and W. Nazarewicz,
Phys. Rev. C {\bf 65}, 054322 (2002).
\bibitem{pmr08}S. Perez-Martin and L. M. Robledo, Phys. Rev. C {\bf 78},
014304 (2008).
\bibitem{sg03}S. Goriely, M. Samyn, M. Bender, and J. M. Pearson,
Phys. Rev. C  \textbf{68}, 054325 (2003).
\bibitem{col04}G. Col\`o, N. Van Giai, J. Meyer, K. Bennaceur,  and 
P. Bonche, Phys. Rev. C {\bf 70}, 024307 (2004).
\bibitem{wir88} R.B. Wiringa, V. Fiks, and A. Fabrocini, Phys. Rev. C {\bf 38}, 
1010 (1988).
\bibitem{bald08}M. Baldo, P. Schuck, and X. Vi\~nas,
Phys. Lett. {\bf B663}, 390 (2008).
\bibitem{frdm} P.~M\"{o}ller, J. R. Nix, W.D. Myers, and W.J. Swiatecki,
At. Data Nucl. Data Tables {\bf 59}, 185 (1995).
\bibitem{ang04}I. Angeli, At. Data and Nucl. Data Tables {\bf 87}, 185 (2004).
\bibitem{gg09}S. Goriely, S. Hilaire, M. Girod, and S. P\'eru,
Phys. Rev. Lett. {\bf 102}, 242501 (2009).
\bibitem{ber75}B. L. Berman and S. C. Fultz, 
Rev. Mod. Phys. {\bf 47}, 713 (1975).
\bibitem{les06} T. Lesinski, K. Bennaceur, T. Duguet, and J. Meyer, 
Phys. Rev. C {\bf 74}, 044315 (2006).
\bibitem{lns} L.G. Cao, U. Lombardo, C.W. Shen, and Nguyen Van Giai,
Phys. Rev. C {\bf 73}, 014313 (2006).
\bibitem{vb72}D. Vautherin and D. M. Brink, Phys. Rev. C {\bf 5}, 626 (1972).
\bibitem{fpt01}M.~Farine, J.~M.~Pearson, and F.~Tondeur,
Nucl. Phys. {\bf A696}, 396 (2001).
\bibitem{zhou04}X.R. Zhou, G.F. Burgio, U. Lombardo, H.-J. Schulze and W. Zuo,
Phys. Rev C {\bf 69}, 018801 (2004).
\bibitem{lom08} U.Lombardo, private communication. 
\bibitem{baldo10} M. Baldo, private communication.
\bibitem{ste06}A. W. Steiner, Phys. Rev. C {\bf 74}, 045808 (2006).
\bibitem{dan02}P. Danielewicz, R. Lacey, and W. G. Lynch,
Science {\bf 298}, 1592 (2002).
\bibitem{ks06}P. G. Krastev and F. Sammarruca, 
Phys. Rev. C {\bf 74}, 025808 (2006).
\bibitem{fpr78}M. Farine, J. M. Pearson, and B. Rouben,
Nucl. Phys. {\bf A304}, 317 (1978).
\bibitem{ton84}F. Tondeur, M. Brack, M. Farine, and J. M. Pearson,
Nucl. Phys. {\bf A420}, 297 (1984).
\bibitem{rei99}P.-G. Reinhard, Nucl. Phys. {\bf A649}, 305c (1999).
\bibitem{war09}M. Warda, X.Vi\~{n}as, X. Roca-Maza, and M. Centelles,
Phys. Rev. C  \textbf{80}, 024316 (2009).
\bibitem{rei10}P.-G. Reinhard and W. Nazarewicz, arXiv:1002.4140
\bibitem{tsa09}M. B. Tsang, Y. Zhang, P. Danielewicz, M. Famiano, Z. Li,
W. G. Lynch, and A. W. Steiner, Phys. Rev. Lett. {\bf 102}, 122701 (2009).
\bibitem{ckl05}L.-W. Chen, C. M. Ko, and B.-A. Li, 
Phys. Rev. C {\bf 72}, 064309 (2005).
\bibitem{fam06}M. A. Famiano, T. Liu, W. G. Lynch, M. Mocko, A. M. Rogers,
M. B. Tsang, M. S. Wallace, R. J. Charity, S. Komarov, D. G. Sarantites, 
L. G. Sobotka, and G. Verde, Phys. Rev. Lett. {\bf 97}, 052701 (2006).
\bibitem{bab00}B.~A.~Brown, Phys. Rev. Lett. {\bf 85}, 5296 (2000).
\bibitem{tb01}S.~Typel and B.~A.~Brown, Phys. Rev. C {\bf 64}, 027302 (2001).
\bibitem{dm69}W. D. Myers and W. J. Swiatecki, Ann. Phys. (NY), {\bf 55},
395 (1969).
\bibitem{fcp80}M. Farine, J. C\^ot\'e, and J. M. Pearson, Nucl. Phys.
{\bf A338}, 86 (1980).
\bibitem{fcp81}M. Farine, J. C\^ot\'e, and J. M. Pearson, Phys. Rev. C
{\bf 24}, 303 (1981).
\bibitem{jas04}J. Jastrzebski, A. Trzcinska, P. Lubinski, B. Klos,
F. J. Hartmann, T. von Egidy, and S. Wycech,
Int. J. Mod. Phys. E {\bf 13}, 343 (2004).
\bibitem{hof80}G. W. Hoffmann, L. Ray, M. Barlett, J. McGill,
G. S. Adams, G. J. Igo, F. Irom, A. T. M. Wang, C. A. Whitten,Jr.,
R. L. Boudrie, J. F. Amann, C. Glashausser, N. M. Hintz, G. S. Kyle,
and G. S. Blanpied, Phys. Rev. C {\bf 21}, 1488 (1980).
\bibitem{bab07}B.~A.~Brown, G. Shen, G. C. Hillhouse, J. Meng, and
A. Trzci\'nska, Phys. Rev. C {\bf 76}, 034305 (2007).
\bibitem{prex}R. Michaels, P. A. Souder, and G. M. Urciouli,\\
http://hallaweb.jlab.org/parity/prex/pbup2005.pdf.
\bibitem{vid09}I.~Vida\~{n}a, C. Provid\^encia, A. Polls, and A. Rios,
Phys. Rev. C  \textbf{80}, 045806 (2009).
\bibitem{mac89}R. Machleidt, Adv. Nucl. Phys. {\bf 19}, 189 (1989).
\bibitem{chenetal}L.-W. Chen, B.-J. Cai, C. M. Ko, B.-A. Li, C. Shen, and 
J. Xu, Phys. Rev. C {\bf 80}, 014322 (2009).
\bibitem{lietal}T. Li, U. Garg, Y. Liu, R. Marks, B. K. Nayak, P. V. Madhusudhana Rao,
M. Fujiwara, H. Hashimoto, K. Nakanishi, S. Okomura, M. Yosoi,
M. Ichikawa, M. Itoh, R. Matsuo, T. Terazono, M. Uchida, Y. Iwao, T. Kawabata,
T. Murakami, H. Sakaguchi, S. Terashima, Y. Yasuda, J. Zenihiro, H. Akimune, 
K. Kawase, and M. N. Harakeh, Phys. Rev. C {\bf 81}, 034309 (2010).
\bibitem{bla80} J.P. Blaizot, Phys. Rep. {\bf 64}, 171 (1980).
\bibitem{piek09a}J. Piekarewicz and M. Centelles, Phys. Rev. C{\bf 79}, 054311
(2009).
\bibitem{piek09b}J. Piekarewicz, J. Phys. G {\bf 37}, 064038 (2010). 
\bibitem{pea10}J. M. Pearson, N. Chamel, and S. Goriely, 
Phys. Rev. C {\bf 82}, 037301 (2010). 
\bibitem{col09}G. Col\`o, {\it Bulk Nuclear Properties}, 
Conference Proceedings (AIP, Melville, NY, 2009), Vol. 1128, p. 59.
\bibitem{jj80}B. K. Jennings and A. D. Jackson, Phys. Rep. {\bf 66}, 
141 (1980).
\bibitem{bla95}J.P. Blaizot, J.F. Berger, J. Decharg\'e, and M. Girod,
Nucl. Phys. {\bf A591}, 435 (1995).
\bibitem{piek07}J. Piekarewicz, Phys. Rev. C {\bf 76}, 031301(R) (2007).
\bibitem{piek05} B.~G.~Todd-Rutel and J. Piekarewicz,
Phys. Rev. Lett. {\bf 95}, 122501 (2005).
\bibitem{zuo03} W. Zuo, C. Shen, U. Lombardo, Phys. Rev. C {\bf 67}, 037301
(2003).
\bibitem{bor84} I.N. Borzov, S.V. Tolokonnikov and S. A. Fayans, 
Sov. J. Nucl. Phys. {\bf 40}, 732 (1984).
\bibitem{frim79} B.L. Friman, and A.K. Dhar, Phys. Lett. B {\bf 85}, 1 (1979). 
\bibitem{she03} C. Shen, U. Lombardo, N. Van Giai, and W. Zuo, 
Phys. Rev. C {\bf 68}, 055802 (2003).
\bibitem{sfb03}A. Schwenk, B. Friman, and G. E. Brown,
Nucl. Phys. {\bf A713}, 191 (2003).
\bibitem{hae07} P. Haensel, A.Y. Potekhin, and D.G. Yakovlev, {\it Neutron stars 1: 
equation of state and structure} (Springer, Berlin, 2007).
\bibitem{wen09}D.-H. Wen, B.-A. Li, and L.-W. Chen,
Phys. Rev. Lett. {\bf 103}, 211102 (2009).
\bibitem{lat91}J. M. Lattimer, C. J. Pethick, M. Prakash, and P. Haensel, 
Phys. Rev. Lett. {\bf 66}, 2701 (1991).
\bibitem{kapl04} D.L. Kaplan, D. A. Frail, B.M. Gaensler, E. V. Gotthelf, S.R. 
Kulkarni, P.O. Slane, and A. Nechita,
Astrophys. J. Suppl. Series {\bf 153}, 269 (2004).
\bibitem{kapl06} D.L. Kaplan, B.M. Gaensler, S.R. Kulkarni, and P.O. Slane,
Astrophys. J. Suppl. Series {\bf 163}, 344 (2006).
\bibitem{sy08}P.S. Shternin and D.G. Yakovlev, 
Astron. Lett. {\bf 34},675 (2008).
\bibitem{pa09} D. Page, J. M. Lattimer, M. Prakash, and A. W. Steiner,
Astrophys. J. {\bf 707}, 1131 (2009).
\bibitem{sg06}S. Goriely, M. Samyn, and J.M. Pearson,
Nucl. Phys. {\bf A773}, 279 (2006).
\bibitem{hg06}S. Hilaire, and S. Goriely, Nucl. Phys. {\bf A779} 63 (2006).

\end{thebibliography}
\end{document}